\documentclass[a4paper, 11pt, oneside]{article}
\usepackage{aapreprint}

\usepackage{indentfirst}
\usepackage{graphicx,wrapfig,float,slashed}
\usepackage{amsmath,amssymb,dsfont,epsfig,graphicx,xcolor}
\usepackage{ragged2e}
\usepackage{epstopdf}
\usepackage[table]{xcolor}
\usepackage{tabu}
\usepackage[toc]{appendix}
\usepackage[normalem]{ulem}
\usepackage[font=small,skip=1pt]{caption}
\allowdisplaybreaks
\usepackage{pifont}
\usepackage{subcaption}

\newcommand{\nc}{\newcommand}
\nc{\non}{\nonumber}
\nc{\hc}{\hbox {H.c.}}
\nc{\noi}{\noindent}
\nc{\barx}{\bar{x}}
\nc{\pbarn}{\;\hbox {pb}}
\nc{\fbarn}{\;\hbox {fb}}

\nc{\hsp}{\hspace{0.5cm}}
\nc{\lsp}{\hspace{1cm}}
\nc{\Lsp}{\hspace{2cm}}
\nc{\LLsp}{\lsp\lsp}
\nc{\lra}{\longrightarrow}
\nc{\p}{\prime}
\nc{\sgn}{\text{sgn}}
\nc{\ph}{\varphi}
\nc{\op}{{\cal O}}
\nc{\eq}{\text{Eq.~}}

\nc{\beq}{\begin{equation}}  \nc{\eeq}{\end{equation}}
\nc{\bea}{\begin{eqnarray}}  \nc{\eea}{\end{eqnarray}}
\nc{\baa}{\begin{array}}     \nc{\eaa}{\end{array}}
\nc{\bit}{\begin{itemize}}   \nc{\eit}{\end{itemize}}
\nc{\ben}{\begin{enumerate}} \nc{\een}{\end{enumerate}}
\nc{\bce}{\begin{center}}    \nc{\ece}{\end{center}}
\nc{\bpm}{\begin{pmatrix}}   \nc{\epm}{\end{pmatrix}}
\nc{\bvt}{\begin{verbatim}}  \nc{\evt}{\end{verbatim}}

\def\lsim{\mathrel{\raise.3ex\hbox{$<$\kern-.75em\lower1ex\hbox{$\sim$}}}}
\def\gsim{\mathrel{\raise.3ex\hbox{$>$\kern-.75em\lower1ex\hbox{$\sim$}}}}

\def\udots{\mathinner{\mkern1mu\raise1pt\vbox{\kern7pt\hbox{.}}\mkern2mu\raise4pt\hbox{.}\mkern2mu\raise7pt\hbox{.}\mkern1mu}}

\def\rts{\sqrt s}

\def\gev{\;\hbox{GeV}}
\def\tev{\;\hbox{TeV}}

\def\mpl{M_{\text{Pl}}}
\definecolor{agray}{rgb}{0.95, 0.95, 0.99}
\def\mphi{m_\phi}
\def\mh{m_h}
\def\lphi{\Lambda_\phi}
\def\gam{\gamma}

\def\mgone{m_1^g}
\def\br{{\rm BR}}

\def\sig{\sigma}
\def\ie{{\it i.e.}}
\def\eg{{\it e.g.}}
\def\fb{~{\rm fb}}

\definecolor{cred}{rgb}{0.6, 0.0, 0.0}

\title{\textcolor{cred}{\huge Implications of the absence of high-mass radion signals}}
\author[a]{Aqeel Ahmed,}
\author[b]{Barry~M. Dillon,}
\author[a]{Bohdan Grzadkowski,}
\author[c]{John~F.~Gunion,}
\author[d]{and Yun Jiang}
\affiliation[a]{Faculty of Physics,
University of Warsaw,
Pasteura 5, 02-093 Warsaw, Poland}
\affiliation[b]{Department of Physics and Astronomy, University of Sussex,
BN1 9QH Brighton, U.K.}
\affiliation[c]{Department of Physics, University of California,
Davis, CA 95616, U.S.A.}
\affiliation[d]{NBIA and Discovery Center, Niels Bohr Institute, University of Copenhagen,\\
Blegdamsvej 17, DK-2100, Copenhagen, Denmark}
\emailAdd{aqeel.ahmed@fuw.edu.pl}
\emailAdd{b.dillon@sussex.ac.uk}
\emailAdd{bohdan.grzadkowski@fuw.edu.pl}
\emailAdd{gunion@physics.ucdavis.edu}
\emailAdd{yunjiang@nbi.ku.dk}

\abstract{{\large
Given the disappearance of the $750\gev $ di-photon LHC signal and the absence of signals at high mass in this and other channels, significant constraints on the mixed Higgs-radion  of the five-dimensional Randall-Sundrum model arise.  By combining all channels, these constraints place a significant radion-mass-dependent lower bound on the radion vacuum expectation value that is fairly independent of the amount of Higgs radion mixing.}
}

\begin{document}

\maketitle
\flushbottom
\section{Introduction}
\label{Introduction}

The most recent ATLAS and CMS data place an upper bound on the cross section for production of a heavy spin-0 resonance as a function of mass in a variety of channels, including $\gam\gam,~Z\gam, t\bar t,~hh,~WW,~ZZ$ and di-jet final states. In this report, we show that these upper bounds imply significant constraints on the  mixed {\it Higgs-radion} sector of the Randall-Sundrum (RS) model \cite{Randall:1999ee}, when choosing the Higgs-like mass eigenstate $h$ to have $\mh=125\gev$ and considering the radion-like mass eigenstate to have $\mphi\geq 300\gev$. The most important bounds near the conformal limit are those from the $\gam\gam$ final state whereas when significantly away from the conformal limit the strongest bounds arise from the $ZZ$ final states.  

We recall that the RS model  is very attractive in that there is no hierarchy problem and economical in the sense that no non-SM extra matter is present beyond that implied by the extra-dimensional context.  The possible importance of a $\phi\to\gam\gam$ signal was first noted in \cite{Giudice:2000av} with more quantitative discussion in Refs.~\cite{Dominici:2002jv,Dominici:2002np,Toharia:2008tm} and \cite{Grzadkowski:2012ng}, the latter showing the dominance of the $\phi \to\gam\gam,gg$ decay modes near the conformal limit in the case of $\mphi=500\gev$. Aside from the earlier version of this paper, the radion/dilaton interpretation of the observed $750\gev$ di-photon excess was also discussed in \cite{Franceschini:2015kwy,Ellis:2015oso, Bellazzini:2015nxw,Cox:2015ckc,Megias:2015ory}.

The RS model consists of one extra spatial dimension bounded by two D3-branes; this can be viewed either as a finite interval or as an $S_1/Z_2$ orbifold. In the limit of no back-reaction, the bulk background geometry is the anti-de Sitter (AdS) space and the spectrum contains a massless radion.  It has been shown that the Goldberger-Wise mechanism is successful both in stabilizing the size of the inter-brane distance in this model and in giving the radion a mass \cite{Goldberger:1999uk,DeWolfe:1999cp}.  In this work we will consider an effective theory where the radion is massive.   In the notation of Refs.~\cite{Dominici:2002jv,Dominici:2002np,Grzadkowski:2012ng} the five-dimensional (5D) metric has the following form,
\beq
ds^2=e^{-2kb_0|y|}\eta_{\mu\nu}dx^\mu dx^\nu-b_0^2dy^2, \label{rs_metric}
\eeq
where $k$ is the curvature of the 5D geometry, $b_0$ is a length parameter for the 5th dimension that is not determined by the action, and $-1/2\leq y\leq 1/2$. For the model considered here, ${1\over 2} kb_0\sim 35$, as appropriate if the RS model is to constitute a full solution to the hierarchy problem. The fluctuation of the $55$-component associated with $b_0$ is referred to as the radion, $\phi_0(x)$.  

The outline of our paper is as follows. In Sec.~\ref{Realistic Higgs-radion model}  we describe the  specific Higgs-radion mixing model that we consider to be particularly strongly motivated. We present a complete phenomenological analysis of this Higgs-radion model in Sec.~\ref{Pheno Higgs-Radion}, and, in particular, use the most recent experimental data to constrain its crucial parameters for the radion mass range $m_\phi=[300-1000]\gev$ . We conclude this report in Sec.~\ref{Conclusions}. In Appendix~\ref{expboundappendix} we collect all the experimental results from ATLAS and CMS (run-1 and run-2) which are employed in this work.

\section{Realistic Higgs-radion of RS model}
\label{Realistic Higgs-radion model}

The literature on  radion models is extensive, but mainly falls into three model categories -- namely those with: the SM on the IR brane \cite{Csaki:2000zn}; the SM in the bulk but the Higgs {doublet} on the IR brane \cite{Csaki:2007ns,Grzadkowski:2012ng,Toharia:2008tm}; or the full SM (including the Higgs  doublet) in the bulk \cite{Cox:2013rva}. Here, we consider a ``mixed" model in which the Higgs  {doublet}, the $t_{L,R}$ and the $b_L$ are localized on the IR brane while the remaining SM fields (including especially the {gauge bosons}) are in the bulk.  We also assume a local bulk custodial symmetry \cite{Agashe:2003zs} of $SU(2)_L\times SU(2)_R \times U(1)_X$, where the $SU(2)_R\times U(1)_X$ fields are broken to $U(1)_Y$ on the UV brane, such that $Y=T_R^3+X$.  This custodial symmetry allows us to obtain a lower scale for the new physics resonances (KK-modes) without violating bounds coming from the electroweak precision observables (EWPO).

Within the above setup we will allow for a Higgs-gravity coupling, $\xi\mathcal{R}_4H^{\dagger}H$~\cite{Giudice:2000av},  
localized on the IR brane, where  $\xi$ is a dimensionless parameter and $\mathcal{R}_4$ is the four-dimensional (4D) Ricci scalar coming from the induced metric on the IR brane.
This results in the following 4D effective Lagrangian for the scalar sector, (see \cite{Dominici:2002np})
\begin{align}
{\cal L}_{\text{eff}}=\frac12 (\partial_\mu\phi_0)^2-\frac12 m_{\phi_0}^2\phi_0^2 -6\xi \Omega\Box \Omega H^\dag H +|D_\mu H|^2-\Omega^4V(H), \label{eff_lagrangian}
\end{align}
where $\phi_0$ is the (unmixed) radion field and $m_{\phi_0}$ its bare  mass.
Above, $\Omega(\phi_0)\equiv 1-\ell \phi_0/v_0$, where
\beq
\label{eq:elldef}
\ell\equiv {v_0 \over \Lambda_\phi}~,
\eeq 
with $v_0=246\gev$ and $\Lambda_\phi\equiv \sqrt6 \mpl e^{-kb_0/2}$ is the vacuum expectation value (VEV) of the radion field. As we will show below, phenomenological constraints make it difficult to accommodate $\Lambda_\phi\lsim 2.5 \tev$, implying that $\ell$ is limited to $\ell\lsim 1/10$.~\footnote{We note that a positive definite kinetic energy for the radion requires
\begin{equation*}
\frac{1}{12}\left( 1-\sqrt{1+\frac{4}{\ell^2}} \right) \leq \xi \leq \frac{1}{12}\left( 1+\sqrt{1+\frac{4}{\ell^2}} \right), \label{kinetic_constraint}
\end{equation*}
which is satisfied for a broad range $\xi$  (including the conformal limit $\xi=1/6$) for $\ell\lsim 1/10$.}

We note that this setup neglects the effects of the back-reaction due to the stabilization mechanism upon the metric which could, in principle, cause deformations to the $AdS_5$ space (corresponding to explicit breaking of the conformal symmetry).  The concern would then be that the suppression of the radion's coupling to the longitudinal modes of the weak gauge boson near the conformal limit ($\xi=1/6$), as needed for a maximal  branching fraction for decay of the radion to $\gam\gam$, might be reduced. This issue deserves further detailed dedicated study.  For now we note that \cite{DeWolfe:1999cp} and \cite{Csaki:2000zn} have shown that the back-reaction on the metric is indeed negligible in certain cases in the absence of Higgs-radion mixing. For phenomenological purposes, a small shift in the values of $\xi$ corresponding to maximal suppression of the $WW$ and $ZZ$ final states will not significantly affect our results, other final states such as $hh$ and $t\bar t$ being of greater importance.

Proceeding, we next rewrite the above Higgs-radion Lagrangian at the quadratic level as
\begin{align}
{\cal L}^{(2)}_{\text{eff}}=-\frac12 (1+6\xi \ell^2)\phi_0\Box\phi_0-\frac12 m_{\phi_0}^2\phi_0^2 +6\xi \ell h_0\Box \phi_0-\frac12 h_0\Box h_0-\frac12 m_{h_0}^2 h_0^2~, \label{eff_lagrangian}
\end{align}
where $h_0$ is the neutral scalar of the Higgs doublet $H$, and $m_{h_0}\equiv \sqrt{2\lambda}v_0$ is the bare Higgs mass.
In the above Lagrangian the $\xi$ term that mixes the Higgs and the radion can be removed by rotating the scalar fields into the mass eigenstate basis,
\begin{align}
\bpm \phi_0\\ h_0\epm=\bpm -a&& -b\\ ~c&&~d\epm \bpm \phi \\ h\epm,   \label{h-r_mixing}
\end{align}
where 
\beq
a=-\cos \theta/Z,\hsp b=\sin\theta/Z, \hsp c=\sin\theta+t\cos\theta, \hsp d=\cos\theta-t\sin\theta,		\label{abcd}
\eeq
with $t\equiv 6\xi\ell/Z$, $Z^2\equiv 1+6\xi\ell^2(1-6\xi)$ and
\beq
\tan2\theta=\frac{12\xi \ell Z m_{h_0}^2}{m_{\phi_0}^2-m_{h_0}^2(Z^2-36 \xi^2\ell^2)}~.  \label{tan2theta}
\eeq

As noted earlier, the $\gam\gam$ final state is of particular importance for constraining the model when $\xi$ is near the conformal limit of $\xi=1/6$.  Thus, we will be providing some details regarding this limit.  In particular, some expansions will be useful.  For now, given that $\ell\lsim 1/10$ for the $\lphi\gsim 2.5\tev$ range of interest, we note that for  a large range of $\xi$ values, including values near the $\xi=1/6$ conformal limit,
it is legitimate to expand Eq.~\eqref{tan2theta} in powers of $\ell$ and express the result in terms of the physical mass parameters, $m_h$ and $m_\phi$:
\beq
\tan2\theta=\frac{12\xi \ell m_{h_0}^2}{Z(m_{\phi}^2+m_{h}^2-2m_{h_0}^2)}\simeq 12 \xi \ell  \left(\frac{m_h}{m_\phi}\right)^2\left[1-3\xi(1-6\xi) \ell^2\right]+\cdots \,,
\label{tan2theta_phy}
\eeq
where the ellipsis stands for terms which are quite small for $m_\phi\geq 300 \gev$ and $\Lambda_\phi \gsim 2.5 \tev$ \footnote{Although this expansion will be useful in later  discussions, to get precise results we employ the exact computation outlined in \cite{Dominici:2002jv}.}. Note that for $\xi=1/6$ and $\ell\lsim1/10$ one obtains $\theta \simeq \ell (m_h/m_\phi)^2 \lsim 1/60$ for $\mphi\geq 300\gev$.

\begin{figure} [t]
\centering
\includegraphics[width=\textwidth]{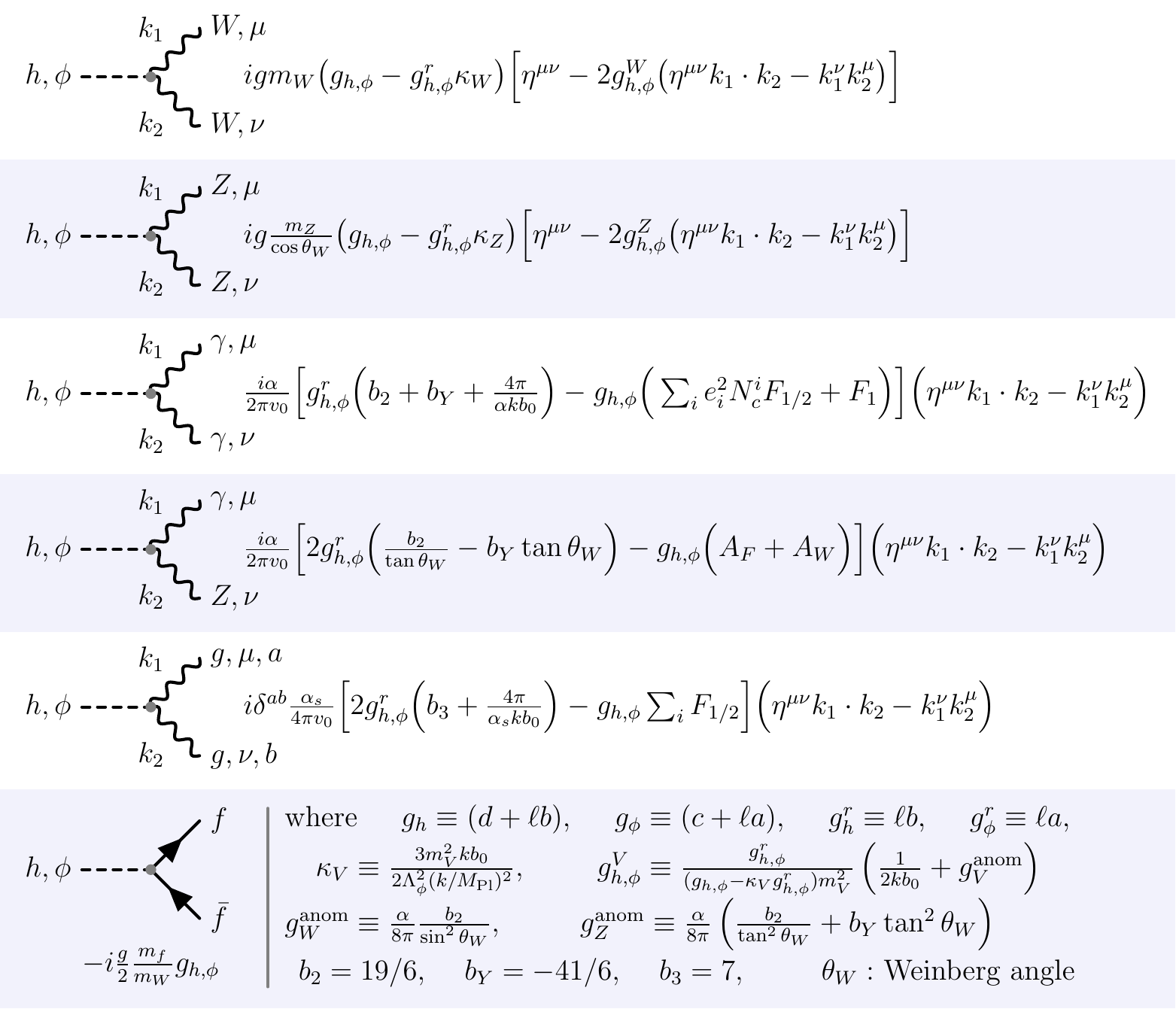}
\caption{Feynman rules for the SM particles couplings with the Higgs $h(x)$ and the radion $\phi(x)$ (the complete list can be found in Refs. \cite{Dominici:2002jv,Grzadkowski:2012ng}). The triangle loop functions $F_{1/2}$, $F_1$, $A_F$ and $A_W$ are given in Ref.~\cite{Gunion:1989we}.}
\label{fig:couplings}
\end{figure}

The most relevant  radion couplings are summarized in Fig.~\ref{fig:couplings}.  There, we have included the ``anomalous" couplings of the radion to  $WW$, $ZZ$ and $Z\gam$ along with the standard ``anomalous'' couplings of the radion to $gg$ and $\gam\gam$. The former ones have been neglected in past studies (see for example Refs.~\cite{Dominici:2002jv}, \cite{Csaki:2007ns} and \cite{Grzadkowski:2012ng}) relative to tree-level couplings proportional to $m_V^2$ (existing for the $ZZ$ and $WW$ cases). For $\xi$ values near the conformal limit, these latter are significantly suppressed and the anomalous couplings may have substantial impact. The
$\kappa_V $ terms and the non-SM tensor structures in the $WW$ and $ZZ$ vertices are both due to the vector bosons being present in the bulk and were first summarized in \cite{Csaki:2007ns}, see also \cite{Grzadkowski:2012ng}. Of course, all these couplings involving the $W$ and $Z$ are obtained  assuming they acquire their masses from the brane-localized Higgs vev. Note that even though the brane contributions to $\phi VV$ are suppressed towards the conformal point (so that anomalous terms might be relevant), there are also large bulk contributions $\propto \kappa_V$ which remain unsuppressed. Therefore one can safely conclude that the anomalous contributions are relevant only for the $\phi gg$, $\phi\gam\gam$ and $\phi Z \gamma $ vertices. 

 For the present paper, the most important point to note from the earlier studies, see e.g. \cite{Giudice:2000av,Dominici:2002jv,Dominici:2002np,Toharia:2008tm,Grzadkowski:2012ng}, is that the coupling of the radion to the trace of the energy momentum tensor leads to the existence of a special choice of $\xi$ where the radion's couplings to SM particles that live on or close to the IR brane are suppressed while the $\phi \gamma\gamma$,  $\phi Z\gam$ and $\phi gg$ couplings have extra ``anomalous" contributions that are not suppressed. As a result, for this special choice of $\xi$  it is possible for the radion to be strongly produced by $gg$ fusion with  primary decay to $gg$ (a di-jet final state) but also with significant branching ratio for decay to $\gamma\gamma$ and  $Z\gam$. Strikingly, the special choice of $\xi$ for which this situation arises when $\mphi\gg\mh$ is close to the conformal limit of $\xi=1/6 \simeq 0.167$. For example, the $t_L$, $t_R$ and $b_L$ (presumed to be localized on the IR brane) have couplings to the radion that are proportional to $g_\phi$ (defined in Fig.~\ref{fig:couplings}) which vanishes at $\xi\sim 0.162$ for small $\ell$ if $\mh=125\gev$ and $\mphi=750\gev$. For larger $\mphi$ values, the value of $\xi $ for which $g_\phi$ vanishes approaches $1/6$ while for $\mphi=300\gev$ it is close to $0.142$. 
In contrast, the $WW,ZZ$ couplings have additional  terms not proportional to $g_\phi$ as a result of their propagation in the bulk. Consequently, as discussed later, the $\phi WW$ and $\phi ZZ$ couplings never actually vanish, but have minimum values at slightly lower values of $\xi$.  The locations of these minima approach the value of $\xi$ where $g_\phi=0$  as the mass of the first KK-gluon excitation, $\mgone$, increases in magnitude.

\section{Phenomenology of the Higgs-radion}
\label{Pheno Higgs-Radion}

\subsection{Bounds from EWPO}

\begin{figure}[t]
\centering
\hspace{-1cm}\includegraphics[scale=0.6]{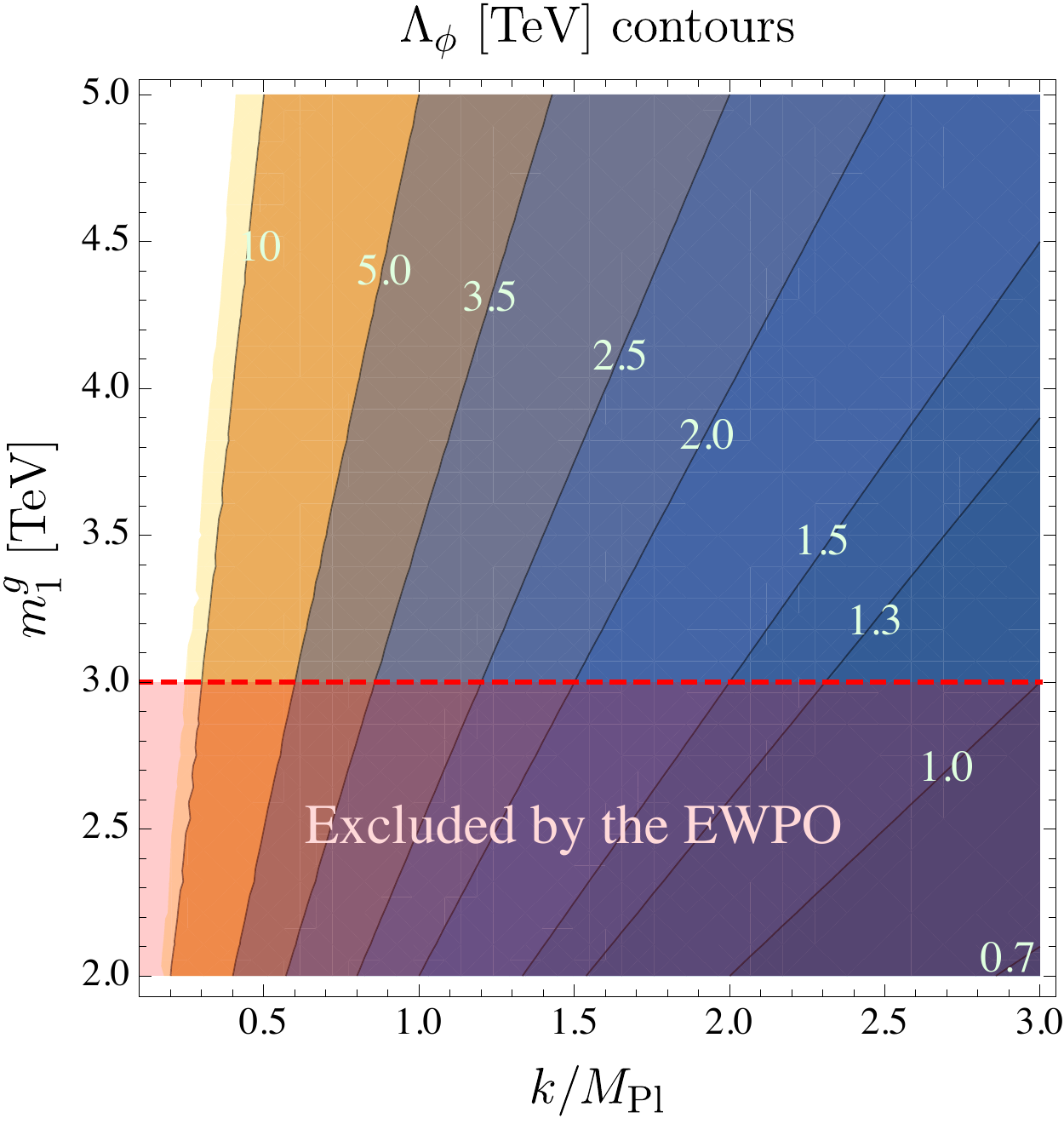}
\caption{Correlation between $m_1^g$ and $k/\mpl$ for different contours of $\Lambda_\phi$ (in the unit of TeV). The region below $m_1^g=3\tev$ (dashed-red line) is excluded by the EWPO.}
\label{figm1glamb}
\end{figure}

Apart from the mixing, the strength of the couplings between the radion and SM fields is controlled by the scale $\Lambda_\phi$, $k/\mpl $ and ${1\over 2}kb_0=\ln(\sqrt6\mpl/\Lambda_\phi)\sim 35$ (see below Eq.~(\ref{rs_metric})).   However, important constraints on the phenomenology of the model arise from the relation of the first KK-gluon mass $\mgone$  to $\Lambda_\phi$ and $k/\mpl$:
\beq
m_1^g=\frac{{x}_1^g}{\sqrt6}\frac{k}{\mpl}\Lambda_\phi, 
\label{m1g}
\eeq
where ${x}_1^g\simeq2.45$ is the 1st zero of an appropriate Bessel function \cite{Dominici:2002jv}. 
Direct bounds from the LHC on $m_1^g$ are typically $\sim 3\tev$~\cite{CMS:lhr,TheATLAScollaboration:2013kha}. Bounds on $\mgone$ from electroweak precision observations (EWPO) are rather model dependent. A value of $\mgone$ as low as $3\tev$ is possible for the model we employ in which the gauge fields (including, especially, the $W$ and $Z$) propagate in the bulk and a bulk $SU(2)_L\times SU(2)_R \times U(1)_X$ symmetry is present.
In Fig.~\ref{figm1glamb} we show the contours of $\Lambda_\phi$ in the $(m_1^g,~k/\mpl)$ plane, 
where the region below $m_1^g=3\tev$ (dashed-red line) is excluded by EWPO. 
The ratio $k/\mpl $ is often assumed to be small. However,  in 
Ref.~\cite{Agashe:2007zd} it is pointed out that the higher curvature terms are suppressed by powers of $R_5/\Lambda^2$ rather than $R_5/M_*^2$, where $\Lambda$ is the scale at which the 5D gravity theory becomes strongly coupled, $M_\ast$ is the 5D Planck mass and $R_5=20 k^2$ is the 5D Ricci scalar. We also recall that the 4D Planck mass $\mpl$ is related to $M_\ast$ by $\mpl^2=M_\ast^2/k$. The crucial point is that $\Lambda$ could be significantly larger than the 5D Planck mass $M_\ast$ while still having a relatively small value for the expansion parameter $R_5/\Lambda^2$. For instance, Naive Dimensional Analysis (NDA) gives an estimate $\Lambda\sim 2\sqrt[3]{3} \pi M_\ast$. With this observation one finds that the ratio $k/M_{\text{Pl}}$ would only have to be $\leq 3$ for $R_5/\Lambda^2<1$, for which there is no need to invoke higher order gravity terms.  

The implications for potential observation of  a Higgs-radion signal at colliders are the following. The signal strengths in the various observable channels decrease with $\lphi^{-2}$ (since the couplings of $\phi$ to possible initial states scale as $\lphi^{-1}$ while the branching ratios for $\phi$ decay to a given final state are roughly independent of $\lphi$).  As we will see, the current absence of signals implies a fairly significant $\mphi$-dependent lower bound on $\lphi$, roughly $\lphi\gsim 2.5\tev$ at large $\mphi$, increasing to larger values at smaller $\mphi$.  
From Fig.~\ref{figm1glamb}, we observe  that for a fixed value of $\lphi$ a larger $\mgone$ requires a larger value of $k/\mpl$ to achieve this particular value of $\lphi$ --- taking the example of $\lphi=2.5\tev$, for $\mgone>3\tev$ we would need $k/\mpl\gsim 1.2$, whereas for $\mgone>5\tev$ $k/\mpl\gsim  2$ is required.  Thus, if the bound on $\mgone$ increases  then the minimum value of $k/\mpl$ must increase in order to achieve a given value of $\lphi$.  Should the LHC eventually set a direct lower bound of $\mgone>10\tev$, and if we take the maximum value allowed for $k/\mpl$ to be 3, then $\lphi$ cannot be smaller than about $3.3\tev$.  We will see that at lower $\mphi\sim 300\gev$, direct searches for a radion signal already exclude such a low value of $\lphi$ whereas at high $\mphi\gsim 500\gev$ values of $\lphi$ somewhat below $3.3\tev$ are still allowed by current data.  Thus, there is a complementarity between LHC searches for the first excited gluon and for a radion signal.

Returning to the bound on $\mgone$ coming from EWPO, we recall that without a bulk custodial symmetry, the $S$ and $T$ parameters both receive large contributions from the mixing of SM gauge bosons with KK gauge bosons \cite{Agashe:2003zs,Cabrer:2011fb,Carena:2002dz,Dillon:2014zea}. These contributions to the $S$ and $T$ parameters decrease in size as $\mgone$ increases. In practice, the strongest constraint arises from the $T$ parameter, yielding a lower bound on $\mgone$ of $\sim 10$~TeV. To weaken this bound, one can first consider imposing a custodial $SU(2)_R$ bulk symmetry. In this case, the contributions to $T$ from KK gauge mixing vanish, but those to $S$ remain unchanged, implying an even larger lower bound on $\mgone$. However, if we break the $SU(2)_R$ symmetry slightly, then the $T$ parameter can be tuned to ``match" the $S$ parameter so as to remain within the 95\% CL LEP $S$--$T$ ellipse provided $\mgone\gsim 3\tev$. It is also important to comment here that the above mentioned custodial symmetry, which keeps the $S$ and $T$-parameters within the experimental bounds, also protects the $Z\bar b b$ vertex from receiving large corrections, see for instance Ref.~\cite{Agashe:2006at}. Finally, corrections to $S$ and $T$-parameters due to $\phi VV$ and $\phi \phi VV$ couplings must be considered. However, these are adequately suppressed so long as current experimental limits on the cross section for radion production and decay to vector bosons are obeyed (as will be the case for our model).  Thus, it is appropriate to explore the Higgs-radion approach for a lower bound on $\mgone$ of order 3--5 TeV.

\subsection{Cross section and branching ratio computations} 

\begin{figure}[t]
\centering
\includegraphics[width=0.33\textwidth]{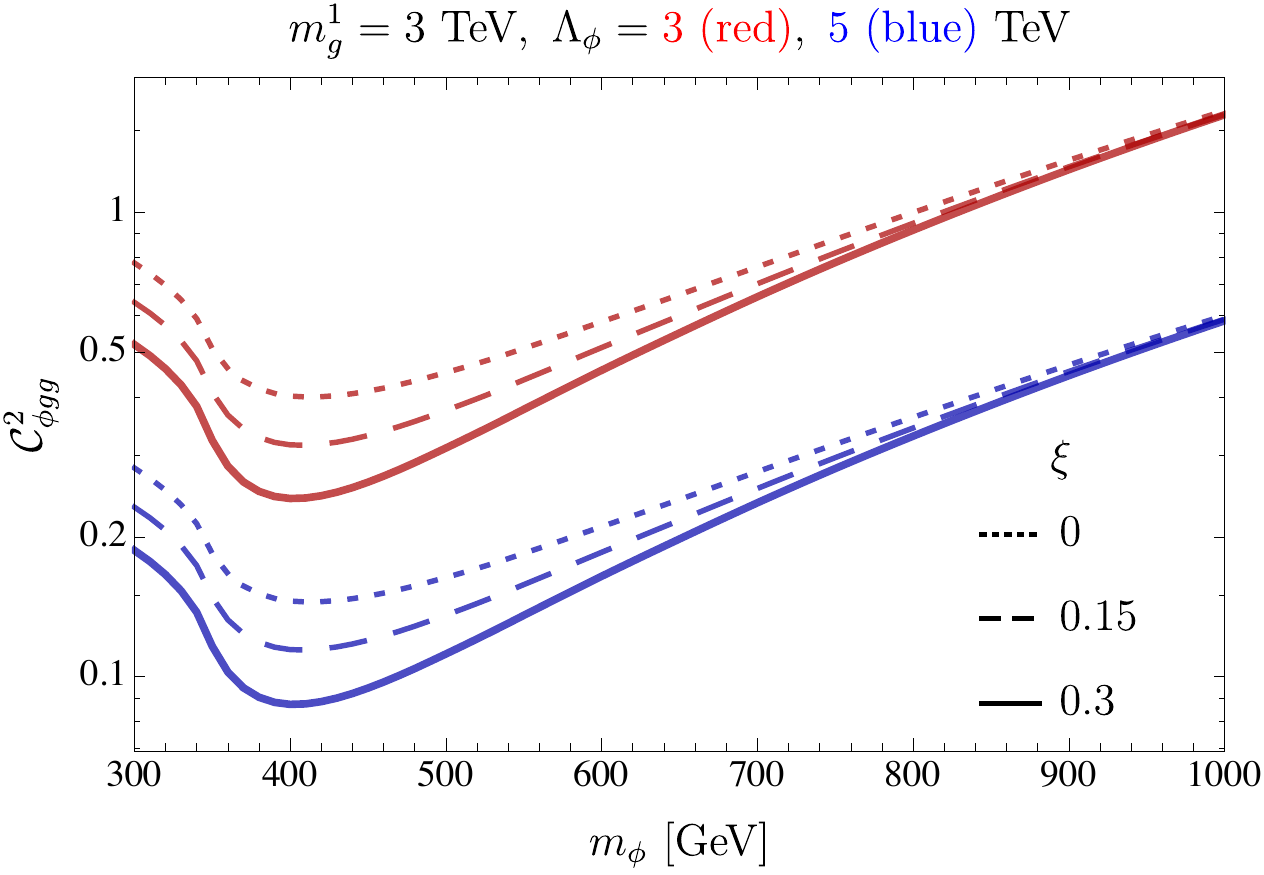}\includegraphics[width=0.33\textwidth]{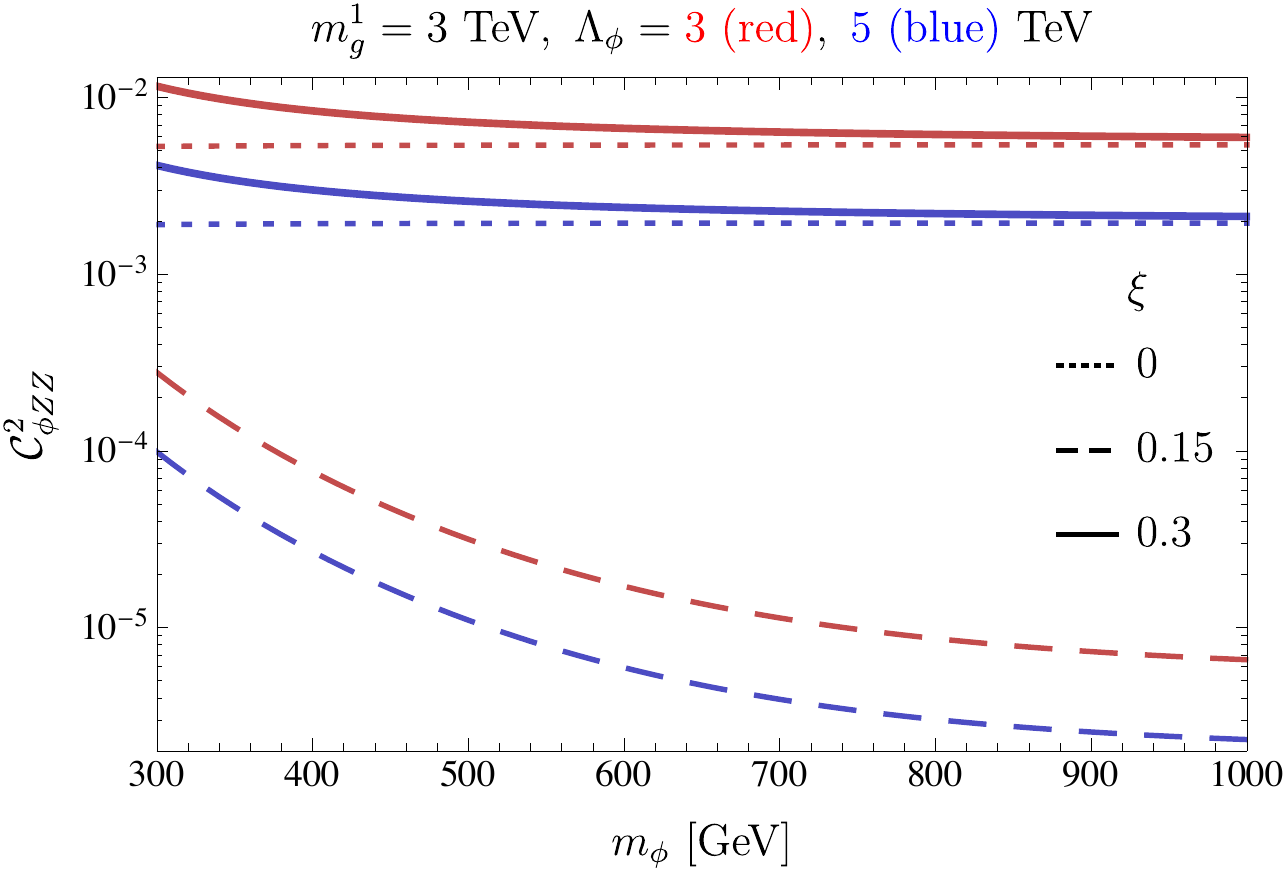}\includegraphics[width=0.33\textwidth]{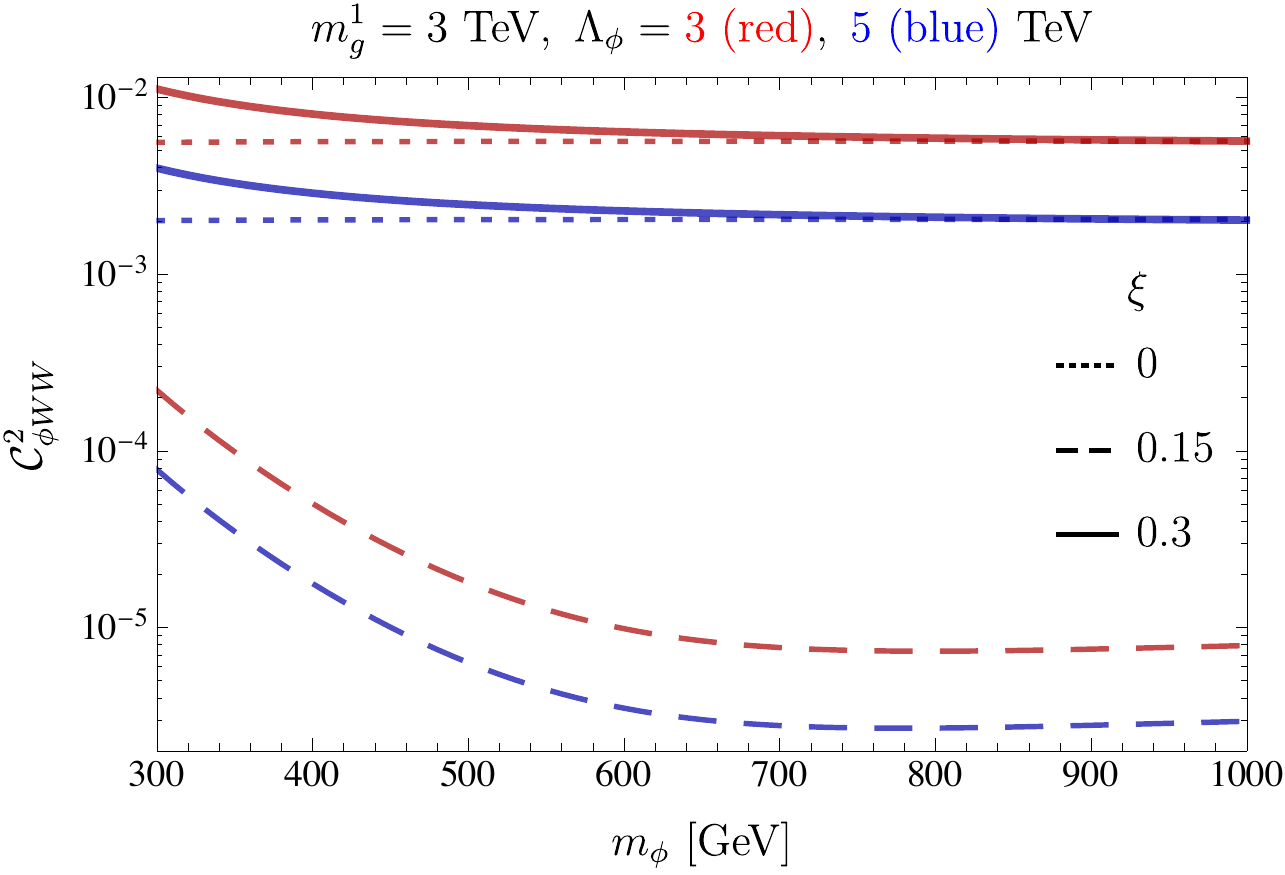}
\caption{These plots show the $m_\phi$ dependence of ${\cal C}_{\phi gg}^2,{\cal C}_{\phi ZZ}^2$, and ${\cal C}_{\phi WW}^2$, the effective couplings of the radion to $gg, ZZ$ and $WW$, respectively, .}
\label{fig:cphi}
\end{figure}

We provide here a few details about our procedures for computing the cross sections and branching ratios for the radion.
The cross-sections for different initial and final states are given by:
\beq
\sigma(YY\to\phi\to XX)=\sigma^{\text{SM}}(YY\to h)|_{m_h\!=\!m_\phi}{\cal C}_{\phi YY}^2  \text{Br}(\phi\to XX),
\eeq
where $\sigma^{\text{SM}}(YY\to h)$ with $YY=gg$ or $VV$ is the Higgs-like scalar production cross-section as obtained by the Higgs Cross Section Working Group (HCWG)~\cite{deFlorian:2016spz} calculated at the radion mass, and the effective couplings ${\cal C}_{\phi YY}$ is defined as,
\beq
{\cal C}^2_{\phi YY} \equiv \frac{\sigma(YY \to \phi)}{\sigma^{\text{SM}}(YY \to h)|_{m_h\!=\!m_\phi}}=\frac{\Gamma(\phi\to YY )}{\Gamma^{\text{SM}}(h\to YY )|_{m_h\!=\!m_\phi}}~,
\eeq
where we obtain the SM partial decay widths $\Gamma^{\text{SM}}(h\to YY )|_{m_h\!=\!m_\phi}$ from HCWG ~\cite{deFlorian:2016spz} and $\Gamma(\phi\to YY )$ is obtained using the Feynman rules given in Fig.~\ref{fig:couplings}. As an important example, for $gg\to \phi$ one finds
\beq
{\cal C}^2_{\phi gg} = 
\left|
g_\phi - 2 g^r_\phi\frac{b_3+\frac{4\pi}{\alpha_s k b_0}}{\sum_iF_{1/2}}
\right|^2~.
\eeq
In Fig.~\ref{fig:cphi} we plot ${\cal C}_{\phi gg}^2$, ${\cal C}_{\phi ZZ}^2$ and ${\cal C}_{\phi WW}^2$, the effective couplings of the radion to  $gg$, $ZZ$ and $WW$, respectively. The first determines the rate for $gg\to\phi$, the main production channel, while the second and third determine the rates for $ZZ\to\phi$ and $WW\to\phi$, respectively.  

\begin{figure}[t]
\centering
\hspace{-2mm}\includegraphics[width=0.33\textwidth]{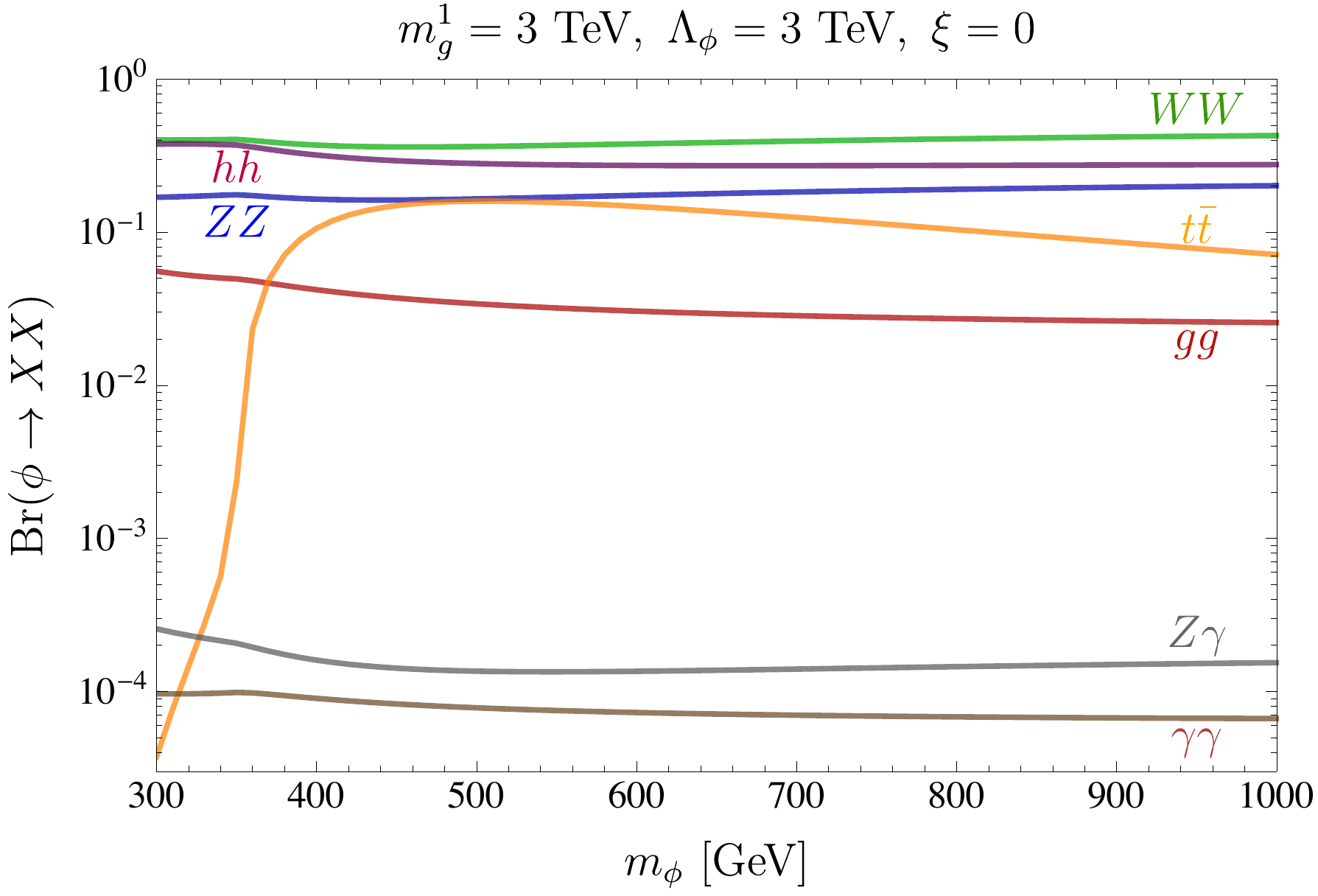}
\hspace{-2mm}\includegraphics[width=0.33\textwidth]{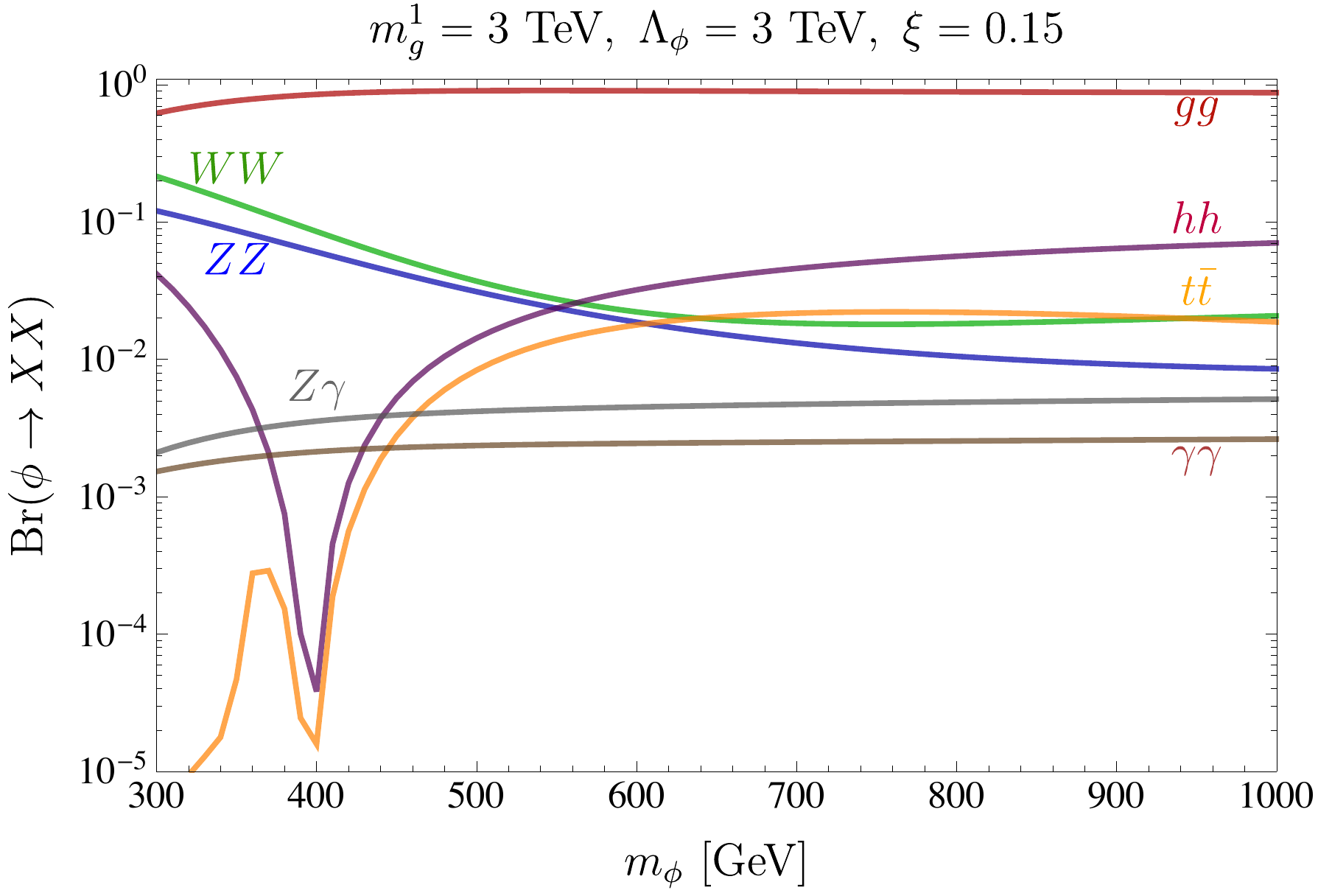}
\hspace{-2mm}\includegraphics[width=0.33\textwidth]{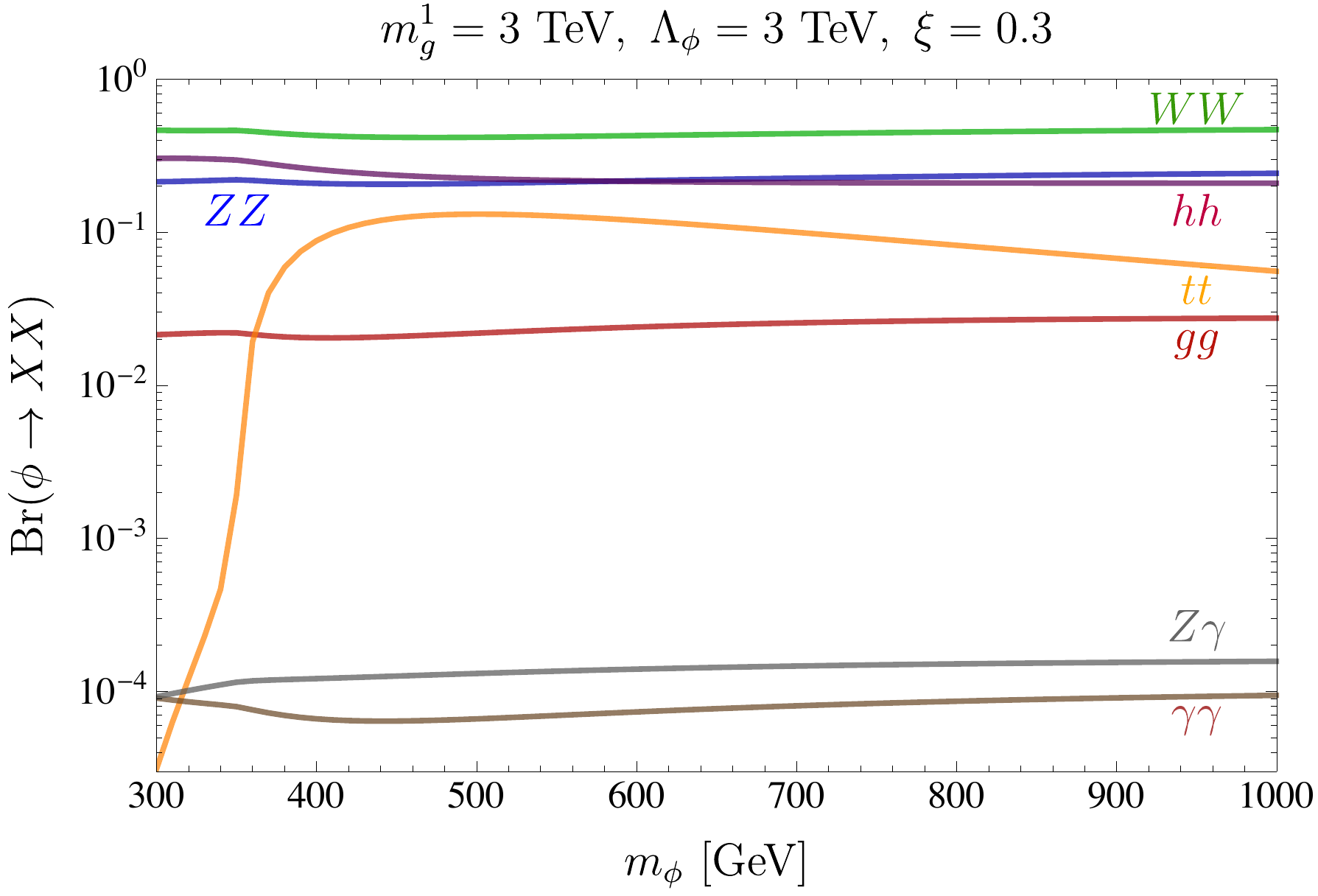}
\caption{These plots show the branching ratios of the radion to different final states as a function of the radion mass. We consider $m_1^g=3\tev$ and plot results for $\xi=0,0.15$ and $0.3$, from left to right, respectively.}
\label{fig:phibrs}
\end{figure}

In Fig.~\ref{fig:phibrs} we illustrate some important branching ratios of the radion for different final states as a function of the radion mass for $\xi=0,~0.15$ and 0.3. For a given production mode (\eg\ gluon fusion), the relative magnitudes of these branching ratios determine the relative rates for the various final states. These plots focus on high radion masses in the interval [300,1000] GeV. Note the dramatic differences between the $\xi=0.15$ results (\ie\ near the conformal limit) and those for $\xi=0$ and $\xi=0.3$.

\subsection {Radion signal in the $\gam\gam$ channel}

\begin{figure}[t]
\centering
\includegraphics[width=0.5\textwidth]{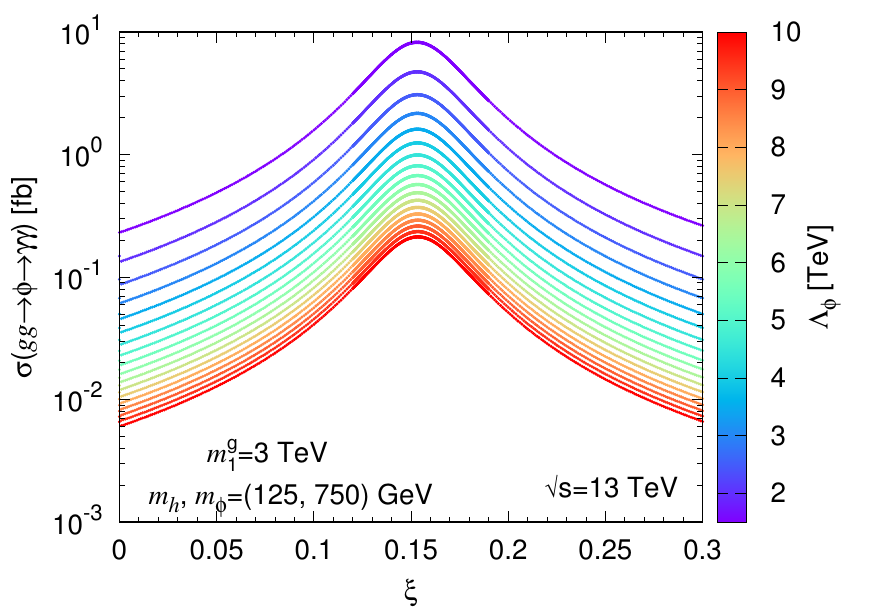}
\hspace{-5mm}
\includegraphics[width=0.495\textwidth]{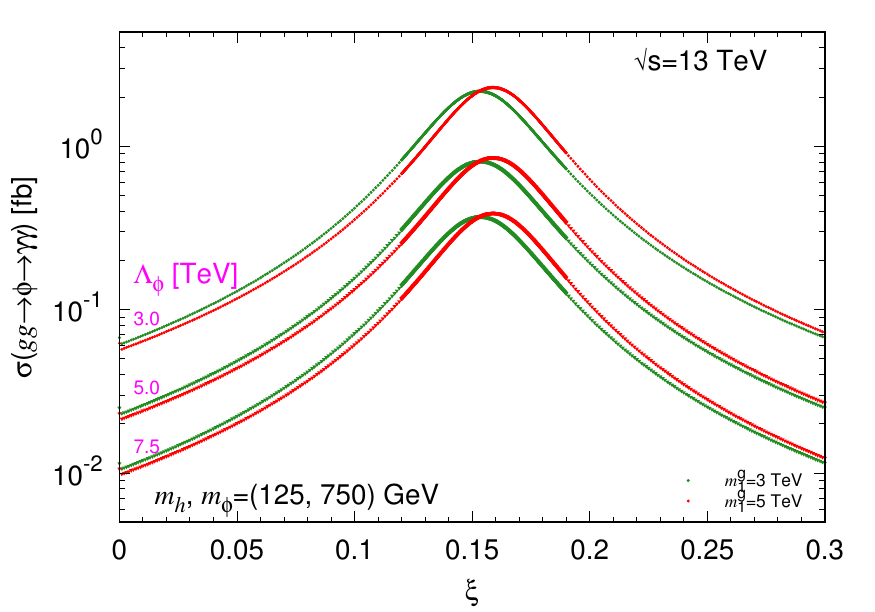}
\caption{The left plot shows the cross-section  $\sigma(gg\to\phi\to \gamma\gamma)$ as a function of $\xi$ for $m_h=125\gev$, $m_\phi=750\gev$ and $m_1^g=3\tev$ with different choices of $\Lambda_\phi$ as indicated by the coloration. The right plot shows  $\sigma(gg\to\phi\to \gamma\gamma)$ for different values of $m_1^g$ and $\lphi$. The accumulation of lines at the red end of the spectrum in the left plot (and many subsequent plots of this type) is simply an artifact of using a linear scale for the coloring vs. a log scale for the cross section.  The $1/\lphi^2$ behavior of the $\gam\gam$ rate (and rates for all other final states) is very apparent in the right figure.}
\label{figphi3tev}
\end{figure}

The $\gam\gam$ channel is quite crucial in that it provides the largest potentially observable radion signal when the mixing parameter $\xi$ is close to the conformal limit $\xi=1/6$. Because the model is particularly attractive near the conformal limit, we will discuss the $\gam\gam$ channel in detail. To understand the nature of the the $\gam\gam$ signal, let us take the example of $\mphi=750\gev$.
We first display in Fig.~\ref{figphi3tev}  the cross section $\sigma(gg\to\phi\to\gamma\gamma)$ as a function of $\xi$. In the left plot we fix $m_1^g=3$ TeV and color-code according to the value of $\Lambda_\phi$ , while in the right plot we show the variation with  $m_1^g$ for  $\Lambda_\phi= 3,$ 5 and 7.5 TeV. Recall that $\mgone\geq 3\tev$ is required for consistency with EWPO and bounds from direct searches \cite{CMS:lhr,TheATLAScollaboration:2013kha}. 
We observe that the maximum  $gg\to \phi\to \gam\gam$ cross section (at $\xi\sim 0.15,~0.16$ for $\mgone=3\tev,~5\tev$ respectively) depends strongly on $\lphi$, but only weakly on $\mgone$.  The current Run-2 upper bounds from ATLAS~\cite{ATLAS:2016eeo} and CMS~\cite{Khachatryan:2016yec} at $750\gev$ are $1.5\fb$ and $2.2\fb$, respectively. We see that at the maximum point, $\lphi\gsim   3\tev$ is implied by the LHC data.\footnote{This is to be compared to the previous ``signals" of about $10\fb$ and $5\fb$~\cite{TheATLAScollaboration:2015mdt,CMS:2015dxe}, respectively, which would have required $\lphi\sim 1.5-2\tev$. } Of course, for $\xi$ values away from the maximum point the $\gam\gam$ cross section limits do not strongly constrain $\lphi$.  However, as we shall soon discuss, for such $\xi$ values the cross sections in other channels are sufficient to constrain $\lphi$ to lie above about $3\tev$.

The fact that the $gg\to\phi \to\gam\gam$ cross section is maximal near the conformal limit is entirely due to the suppression of all $\phi$ couplings other than those to $gg,\gam\gam,Z\gam$. Note that in the region $\xi\in [0, 0.3]$ the radion couplings to $\gamma\gamma$, $gg$  and $Z\gam$ are dominated by the contribution proportional to $g^r_\phi \sim -\ell + \op(\ell^3)$ which depends on $\xi$ very weakly, through terms $\propto \ell^3 $. The remaining contribution $\propto g_\phi$ is negligible near $\xi=1/6$. Therefore, the resulting $\xi$ dependence of the radion couplings to $\gamma\gamma$ or $gg$ is rather mild and consequently $\Gamma(\phi\to gg)$ (and hence $\sig(gg\to \phi)$) and $\Gamma(\phi\to\gamma\gamma)$ vary slowly for $\xi\in [0, 0.3]$. In contrast, the main contributions to the couplings of all the SM particles (except $gg$, $\gam\gam$ and $Z\gam$) to the $\phi$ actually vanish or have minima in the vicinity of the conformal point, $\xi=1/6$, leading to maximal $\br(\phi\to gg)$ and $\br(\phi\to \gam\gam)$. However, as discussed below, the magnitudes of $\br(\phi\to gg)$ and $\br(\phi\to\gam\gam)$  depend crucially upon whether or not the gauge bosons are confined to the TeV brane or propagate in the bulk.  The di-photon cross section is maximal in the latter case.

\subsection{Radion couplings near the conformal limit}

Since the $\gam\gam$ signal is maximized near the conformal limit, it is of particular interest to analyse the radion couplings in this limit.  We will use $\mphi=750\gev$ for numerical illustration. 
There are four critical $\phi$ couplings:  the coupling to the top and bottom quarks, $\propto g_\phi$; the coupling to $hh$, $g_{\phi hh}$; and the couplings to $WW$ and $ZZ$.
It is helpful to examine these couplings analytically in the limit of small $\ell=v_0/\lphi$. Using the expression $g_\phi=c+\ell a$, the explicit forms for $c$ and $a$ are given earlier in \eqref{abcd}, and the approximations $\tan 2\theta \simeq 2\theta$, $\sin \theta\simeq \theta$, $\cos\theta\sim 1$ as appropriate for very small $\theta$, see Eq.~(\ref{tan2theta_phy}), one finds
\beq
g_\phi= \ell \left[6\xi \left(\frac{\mh}{\mphi}\right)^2+6\xi-1\right] \buildrel{\mphi=750~\rm{GeV}}\over{\simeq} \ell \left({37\over 6}\xi -1\right)\label{g_phi}
\eeq
where we have neglected terms of order $\ell^3$ relative to $\ell$ given that   $\ell=v_0/\lphi$ is a small number. This derives the solution $\xi\simeq 0.162162$ for $g_\phi=0$. 
At this point, the (brane localized) $t\bar t$ and $b\bar b$ couplings of the $\phi$ will vanish.  
The $hh$ coupling takes the form given in Fig.~34 of \cite{Dominici:2002jv}.  Despite the complicated form of the coupling, in the limit of small $\ell$ one finds a relatively simple result,
\beq
{g_{\phi hh}\lphi\over \mphi^2}=(1- 6\xi)+{2m_h^2\over m_\phi^2}(1-9\xi)-18\xi \left({m_h^2\over m_\phi^2}\right)^2 \buildrel{\mphi=750~\rm{GeV}}\over{=}{19\over 18}(1 - 6.17105 \xi)\,, \label{g_phihh}
\eeq
where terms of order $\ell^2$ are neglected.  Numerically, $g_{\phi hh}$
vanishes for $\xi=0.162047$, \ie\ very close to the value for which $g_\phi$ vanishes.  
Finally, we consider the $WW,ZZ$ couplings of the $\phi$.  From Fig.~\ref{fig:couplings} we see that the terms in the $\phi VV$ couplings proportional to $\eta^{\mu\nu}$ (\ie\ of SM-like form) are multiplied by $\eta_V\equiv (g_\phi-g_\phi^r\kappa_V)$. For small $\ell$, we find the form
\beq
\eta_V=g_\phi-g_\phi^r\kappa_V\simeq \ell \left[\kappa_V+6 \xi \left(\frac{m_h}{\mphi}\right)^2+6 \xi-1\right]\,,\label{eta_V}
\eeq
where $\kappa_V=\frac{3k b_0 m_V^2}{2\lphi^2 (k/\mpl)^2}\simeq {105 m_V^2\over {\mgone}^2}$ for $k b_0/2\sim 35$ using the very good approximation $\lphi (k/\mpl)=\mgone$, see Eq.~(\ref{m1g}). For $\mgone=3\tev$, one finds $\kappa_W=0.0761$ and $\kappa_Z=0.0981$. As a result, for the example of $\mphi=750\gev$, $\eta_V$ vanishes at $\xi=0.150$, $0.146$ in the $W$, $Z$ cases, respectively. Of course, the zeroes shift closer to the $\xi=0.162$ point for $\mgone=5\tev$, occurring at $\xi=0.158$, $0.157$, respectively. It is important to note that in above equations \eqref{g_phi}, \eqref{g_phihh} and \eqref{eta_V} all the next to leading order contributions to these couplings are suppressed by at least $\ell^2\leq1/100$ (recall that $\ell\equiv v/ \lphi\leq 1/10$ for $\lphi\geq 2.5\tev$) relative to the terms we have kept in our analytic discussion. 
Of course, in our numerical analysis we considered the exact forms of and relations between all the couplings.

However, there is more to the story. In Fig.~\ref{fig:couplings} one sees terms in the $\phi VV$ couplings coming from off-brane effects proportional to $\eta_V g_\phi^V=g_\phi^r /(2 m_V^2 k b_0)\sim -\ell /(2m_V^2 k b_0)$ which do not vanish for $\eta_V=0$.  However, the $\phi VV$ couplings do  have very small values in this limit.  As a result, the $WW$ and $ZZ$ widths will be minimal quite close to the $\xi$ values for which $\eta_W$ and $\eta_Z$ are zero.  Nonetheless, the fact that the locations of these minima differ from the $\xi\sim 0.162$ location of $g_\phi=0$ and $g_{\phi hh}=0$ means that the $\phi\to \gam\gam$ branching ratio will be maximal at some intermediate $\xi$ value.

In contrast to the above, the $\gamma\gamma$ and $gg$ couplings of the $\phi$ are slowly varying with $\xi$ in this conformal region,
see also Ref. \cite{Giudice:2000av}. Meanwhile, the Higgs coupling coefficient $g_h\simeq \cos\theta\simeq 1+\ell^2(m_h/m_\phi)^4/2+\cdots$,  implying a very SM-like $125\gev$ state given that  $\ell$ and $m_h/\mphi$ are both very small. Note that achieving $g_h=1 $ and $g_\phi=0$ simultaneously is only possible if $\theta=0$ and $\xi=1/6$, which would require $\mphi\to\infty$.  In practice, $\mphi=750\gev$ is not far from this situation.

Another remark concerning the $g_\phi=0$ limit is useful at this point.  From the Feynman rules shown in Fig.~\ref{fig:couplings} 
one sees that if $g_\phi=0$ then $\Gamma(\phi \to gg)$ and $\Gamma(\phi \to \gamma \gamma)$ can be expressed exclusively in terms of the beta function coefficients and the bulk contributions. The ratio of the cross sections for the production of the $\gam\gam$ and $gg$ final states is independent of $\lphi$ in the small $\ell$ limit and for $g_\phi=0$ is given by the ratio of the respective branching ratios
\beq
\frac{BR(\phi\to \gam\gam)}{BR(\phi\to gg)}= \frac{\left[\alpha (b_2+b_Y)+\frac{4\pi}{ kb_0}\right]^2}{N_c \left[\alpha_s b_3+\frac{4\pi}{kb_0}\right]^2} \simeq {1\over {360}}
\label{BRratio}
\eeq
for $N_c=3$, $kb_0=70$ and the low-energy values $\alpha,\alpha_s=1/137,0.12$. However, there are radiative corrections to $\alpha$ and $\alpha_s$ (yielding $\alpha\sim 1/125$ and $\alpha_s\sim 0.1$ at $750\gev$) as well as a $K$ factor of about  $K=1.35$ for the $gg$ final state.  The resulting ratio of cross sections  is accidentally quite close to the $1/360$ number quoted above. 
Note that were the gauge bosons confined to the TeV brane  then the $4\pi/( kb_0)$  terms would be absent and the ratio would be  a factor of roughly 20 smaller,  and the maximum $\gam\gam$ cross section would easily lie below the current ATLAS and CMS bounds even for $\lphi$ as low as $1\tev$. 
Thus, to obtain significant bounds on $\lphi$ (or potential for radion discovery at the LHC) for $\xi$ near the conformal limit it is necessary that the massless gauge bosons, $\gam$ and $g$, propagate in the bulk.  There are several important reasons why this latter is required in the context of the present model.  First, EWPO consistency for $\mgone\sim 3-5\tev$ requires that the $W$ and $Z$ must propagate in the bulk and the underlying $SU(3)\times SU(2)\times U(1)$ group structure then requires that the same must be true for the $\gam$ and $g$.
Second, to naturally accommodate light fermions they must be in the bulk and localized towards the UV brane.  Then in order for these fields to have the correct gauge couplings the SM gauge fields are required to be in the bulk.

\begin{figure}[t]
\centering
\includegraphics[width=0.5\textwidth]{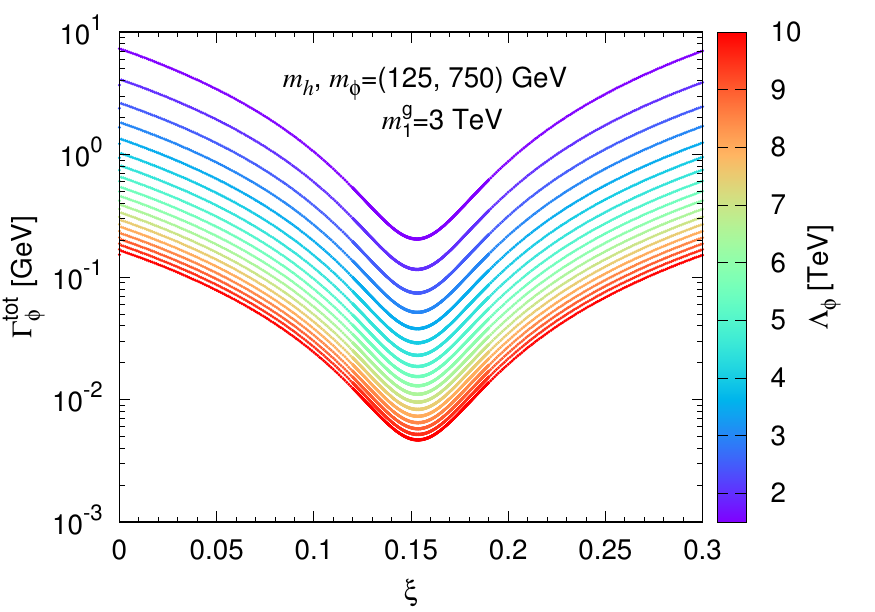}
\hspace{-3mm}
\includegraphics[width=0.5\textwidth]{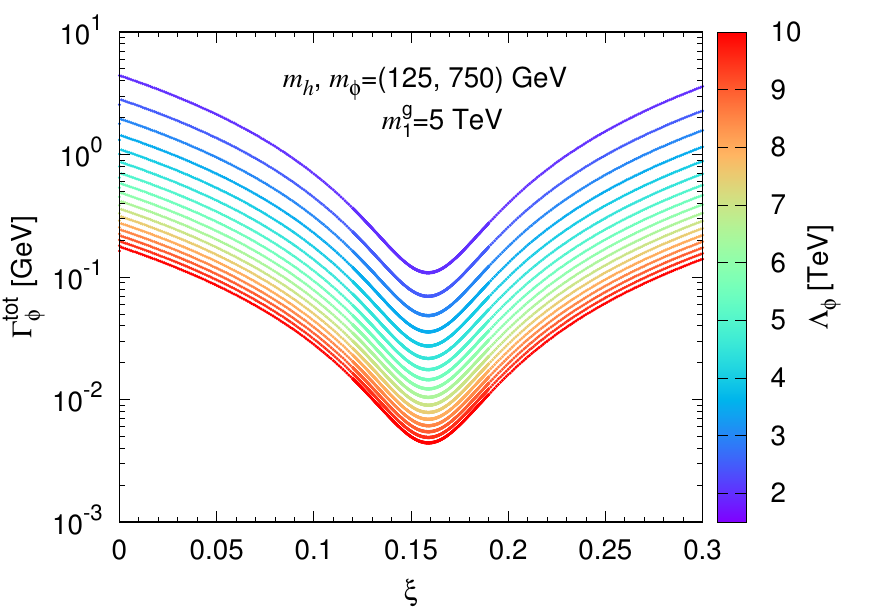}\\
\includegraphics[width=0.5\textwidth]{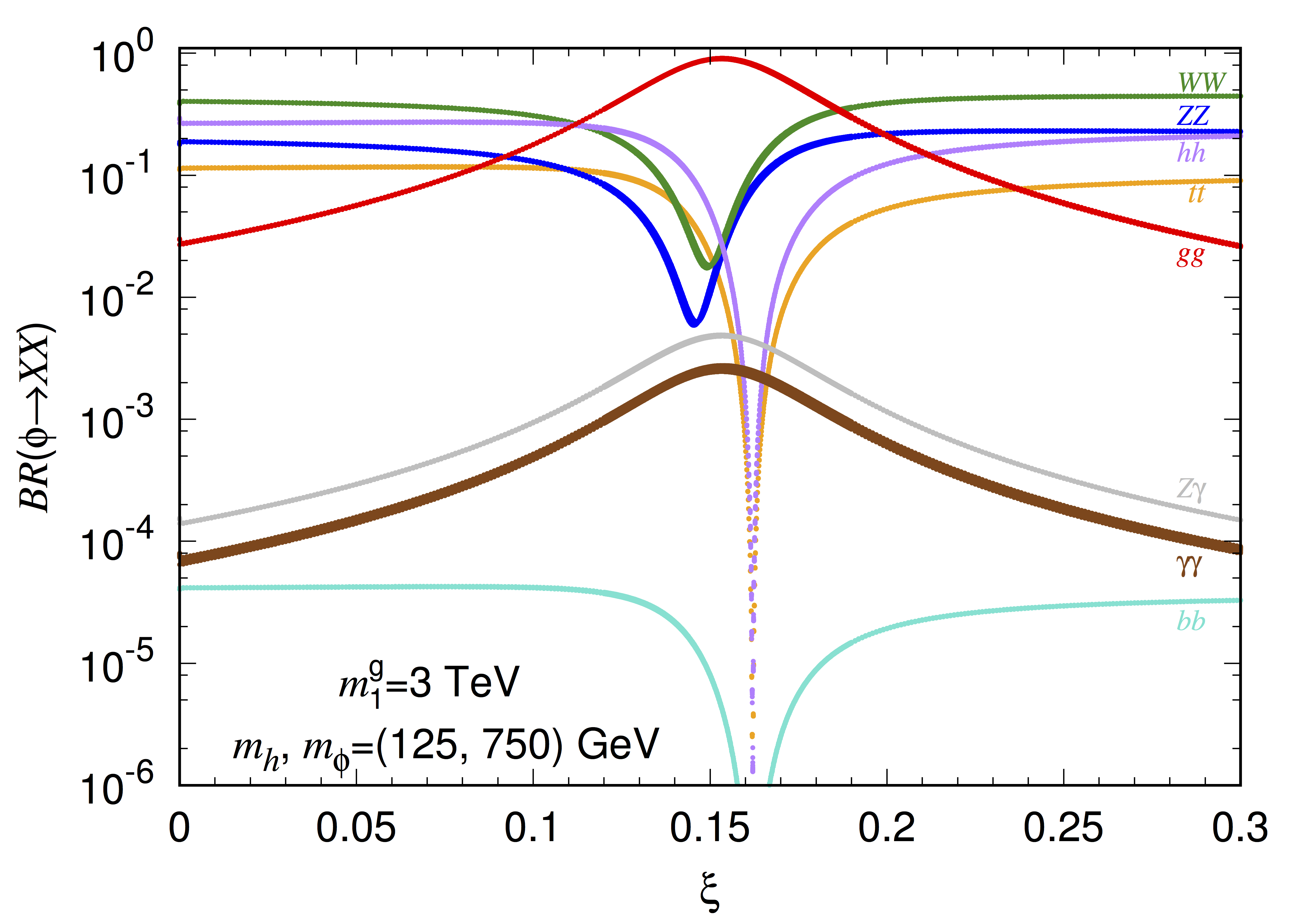}
\hspace{-3mm}
\includegraphics[width=0.5\textwidth]{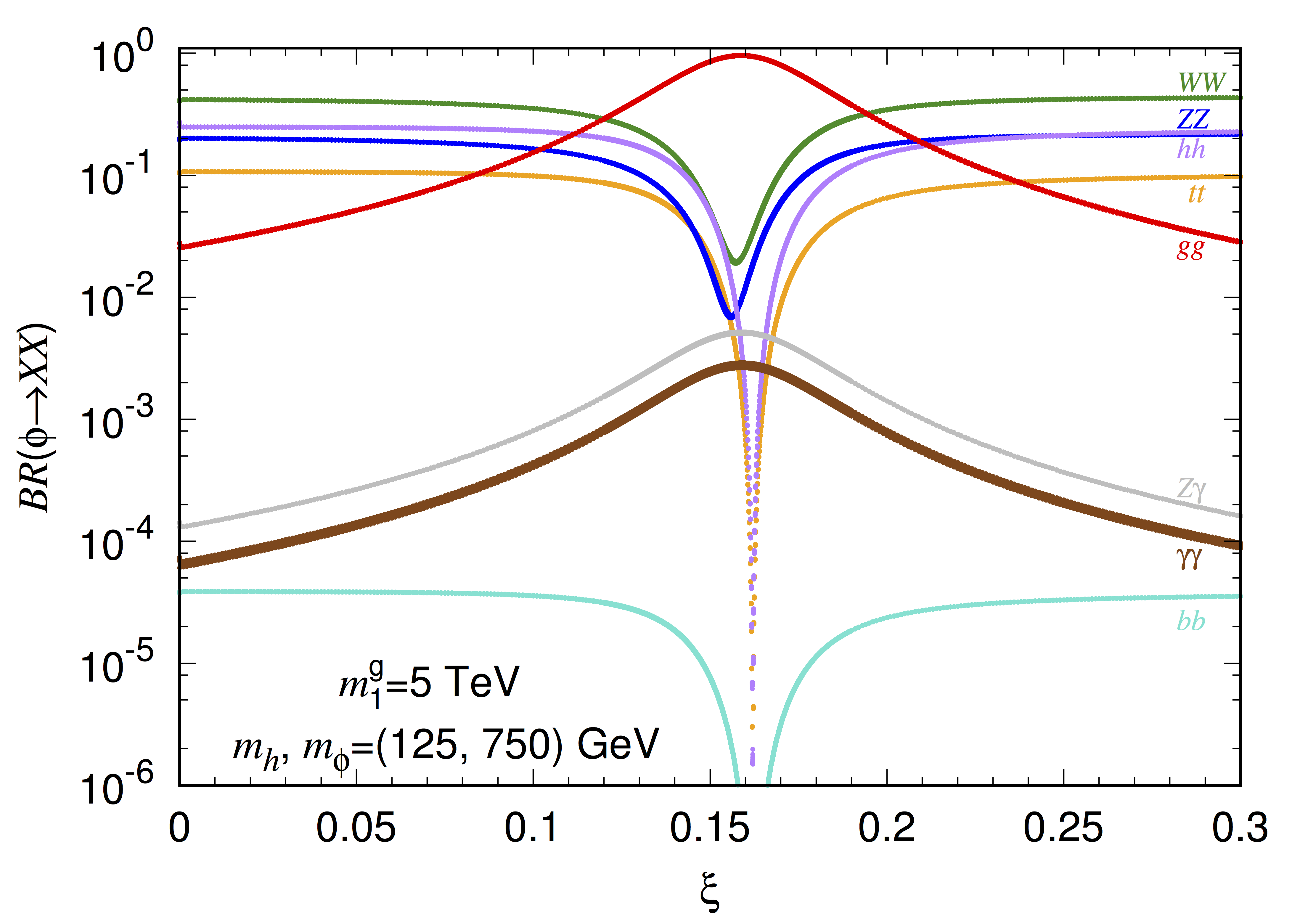}
\caption{The upper graphs show the total width of the radion $\Gamma^{\text{tot}}_\phi$, as a function of $\xi$ for $m_h=125\gev$, $m_\phi=750\gev$, $m_1^g=3\tev$ (left) or $m_1^g=5\tev$ (right)  with different choices of $\Lambda_\phi$. The total width has little dependence on the $m_1^g$ value as expected. The middle and lower graphs show the branching ratios for $\phi\to XX$, $XX=b\bar b, t\bar t, WW,ZZ, hh, gg, \gamma\gamma, Z\gamma$, as functions of $\xi$ for $m_h=125\gev$, $m_\phi=750\gev$ and $m_1^g=3\tev$ (left) or $m_1^g=5\tev$ (right). The width of the curves reflects the variation with $\lphi$ for $\lphi\in[1.5,6]\tev$ in the case of $\mgone=3\tev$ and for  $\lphi\in[1.5,10]\tev$ in the case of $\mgone=5\tev$.  
}
\label{figphixx3tev}
\end{figure}

The phenomenological features for $\phi$ production and decay associated with the above discussion are manifest in the lower panels of Fig.~\ref{figphixx3tev}. There, we show the branching ratios for $\phi$ decay to $XX=b\bar b, t\bar t, ZZ, hh, gg, \gamma\gamma, Z\gamma$, as functions of $\xi$ for $m_h=125\gev$, $m_\phi=750\gev$ and 
$m_1^g=3\tev$ (left) and $5\tev$ (right). In the vicinity of $\xi\simeq 0.162$, one sees that the $gg$ mode dominates radion decay and $\br(\phi\to \gam\gam)\sim \text{few}\times 10^{-3}$, whereas the $t\bar t$, $b\bar b$ and $hh$ branching ratios are negligible. The magnitude of the $WW,ZZ$ branching ratios at $\xi\simeq 0.162$ depends upon $\mgone$. For $\mgone=3\tev$, the dips for the $VV$ final states are significantly separated from the $\xi$ value where $g_\phi=0$ so that one finds $\br(\phi\to WW,ZZ)\sim 0.1$ if $\br(\phi\to t\bar t,b\bar b,hh)=0$. Thus, for $\mgone=3\tev$ and sufficiently low $\lphi$, $gg\to\phi\to VV$ might be observable even if $gg\to \phi \to t\bar t, b\bar b,hh$ are not and vice versa. However, for $m_1^g=5\tev$ the dips in the $VV$ branching ratios are much closer  to the $g_\phi\simeq g_{\phi hh}=0$ point, implying that in the vicinity of this point only limits on the $\gam\gam$ and $Z\gam$ rates would be relevant for constraining $\lphi$. Conversely,  away from the $g_\phi\simeq g_{\phi hh}=0$ point it is the $WW,ZZ,t\bar t, hh$ modes that will provide the strongest limits on $\lphi$ at any given $\mphi$.

In the upper panels of Fig.~\ref{figphixx3tev}, we present the total width of the $\phi$, $\Gamma^{\text{tot}}_\phi$, as a function of $\xi$ taking $m_\phi=750\gev$ and $m_1^g=3\tev$ and $5\tev$. One sees that at the  dip the total width is very small, e.g.  $\leq 0.05\gev$ for $\lphi>2.5\tev$ for $\mgone=3\tev$.
For $\xi$ values  away from the minimal point, $\Gamma^{\text{tot}}_\phi$ grows rapidly, implying that should a signal be seen in the $WW,ZZ,hh,t\bar t$ channels it should be associated with a substantial total width.

A few more concrete comments concerning the conformal point of $\xi\simeq 1/6$ are in order.  
As is well known~\cite{Giudice:2000av,Csaki:2000zn}, if the SM fields are localized on the IR brane, then
the radion couples to the SM as a dilaton
\beq
{\cal L}_{\rm int}^{\phi_0\text{-SM}} = \frac{\phi_0}{\Lambda_\phi} T_\mu^\mu,
\label{radint}
\eeq 
where $T_\mu^\mu$ is the trace of the energy-momentum tensor. If one includes the Higgs-gravity mixing contribution 
in the trace then, using the SM equation of motion for the Higgs field $H$, 
one finds~\cite{Giudice:2000av} in the unitary gauge ($H\to (v+h_0)/\sqrt{2}$) that
\begin{align}
T_\mu^\mu &= -(1-6\xi)\Big[\partial_\mu h_0 \partial^\mu h_0 + m_V^2 V_{\mu}V^{\mu}\Big(1+\frac{h_0}{v_0}\Big)^2
- m_{i}\bar{\psi}_i\psi_i \Big(1+\frac{h_0}{v_0}\Big) - \lambda(v_0+h_0)^4 \Big]\non \\
& -(1-3\xi)m_{h_0}^2(v_0+h_0)^2 +\frac{\alpha}{8\pi}\Big[(b_2+b_Y) F_{\mu\nu}F^{\mu\nu} +2\Big(\frac{b_2}{\tan\theta_W}- b_Y\tan\theta_W \Big) F_{\mu\nu}Z^{\mu\nu}\notag\\
&+\Big(\frac{b_2}{\tan^2\theta_W}+b_Y\tan^2\theta_W \Big) Z_{\mu\nu}Z^{\mu\nu}+\frac{2b_2}{\sin^2\theta_W} W^{+}_{\mu\nu}W^{-\mu\nu} \Big]+\frac{\alpha_s}{8\pi}b_3 \text{Tr} \big[G_{\mu\nu} G^{\mu\nu}\big],
\label{trace}
\end{align}
where $b_3=7$, $b_2=19/6$ and $b_Y=-41/6$ are the $\beta$-function coefficients for QCD, SU(2) and U(1), respectively.
The last two lines above except the first term in the second line are the trace anomaly contributions.
As expected, conformal symmetry would be maintained at the tree level if $\xi=1/6$ (the standard condition
for the scalar-gravity coupled conformal theory) and if no explicit conformal symmetry breaking  is present (requiring absence of a Higgs mass term).
Therefore, if $\xi$ is close to $1/6$, couplings of the $\phi_0$ to the massive SM particles are suppressed either by $(1-6\xi)$ or by the Higgs mass terms, the only non-suppressed couplings being those generated by the trace anomaly.  In our case, only the Higgs field (aside from the $t_L,t_R,b_L)$ is localized on the IR brane.  Of course, the above arguments only hold for the matter fields localized on the IR-brane. All light fermions are localized toward the UV brane in order to have small mass $m_f$, so that their couplings to the radion are suppressed by $m_f/\Lambda_\phi$ and therefore can be neglected. However, for the massive gauge boson fields, the above reasoning is violated by non-neglectable terms generated due to their being in the bulk; these effects are 
encoded in the coefficients $\kappa_V$ in Fig.~\ref{fig:couplings}. This deviation from the conformal limit is illustrated in Fig.~\ref{figphixx3tev} where one can see that the suppression of $BR(\phi\to VV)$ is shifted away from $\xi\sim 1/6$ (the effect being more pronounced for smaller $m^g_1$).

\subsection {Other final states}

We provide an illustrative discussion for the case of $\mphi=750\gev$.
As pointed out in the Introduction, the very large $\br(\phi\to gg)$ seen in Fig.~\ref{figphixx3tev} could lead to  a cross section for $gg\to \phi\to gg$  excluded by the di-jet search at the LHC.  
\begin{figure}[t]
\centering
\includegraphics[scale=0.85]{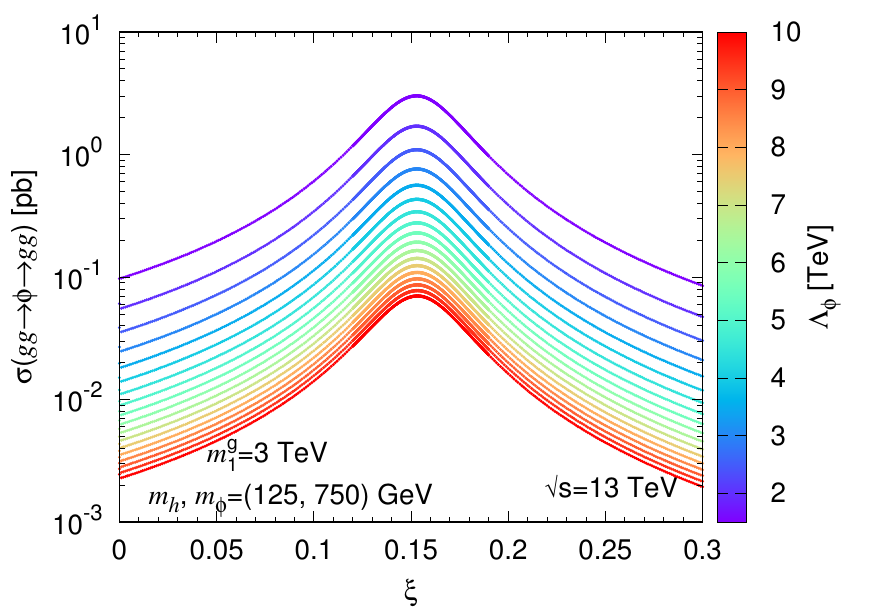}
\caption{We plot $\sigma(gg\to\phi\to gg)$ as function of $\xi$ for $m_h=125\gev$, $m_\phi=750\gev$ and $m_1^g=3\tev$, color-coded by $\Lambda_\phi$.}
\label{figphigg3tev}
\end{figure}
To illustrate the potential for future observation of the $gg$ final state, we plot the $13\tev$ results for $\sigma(gg\to \phi\to gg)$ in Fig.~\ref{figphigg3tev} in a manner analogous to Fig.~\ref{figphi3tev}. For $\xi$ values in the region around the peak of the  di-photon cross section one finds very significant cross sections for lower values of $\lphi$. Of course, one must keep in mind that limits on a di-jet resonance are not as strong as they might be because of difficult QCD backgrounds.  Thus, other final states could (and currently do) yield stronger exclusions.
\begin{figure}[t]
\centering
\begin{tabular}{ccc}
\hspace{-5mm}\includegraphics[scale=0.62]{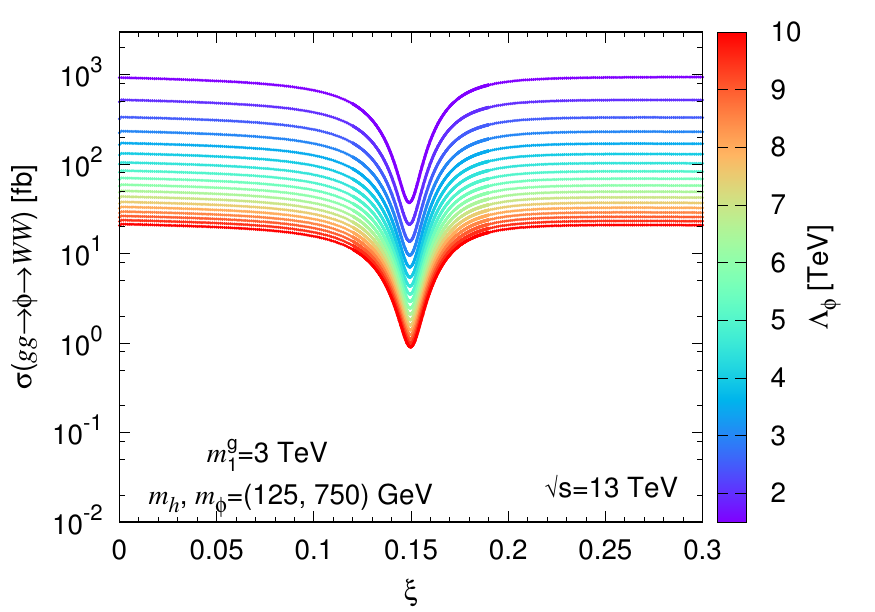}
\hspace{-4mm} \includegraphics[scale=0.62]{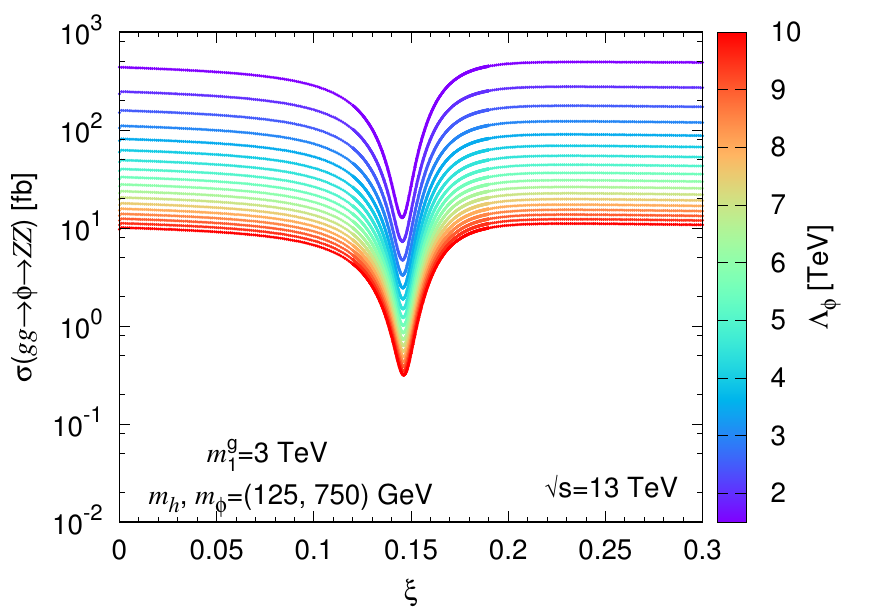}
\hspace{-3mm}\includegraphics[scale=0.62]{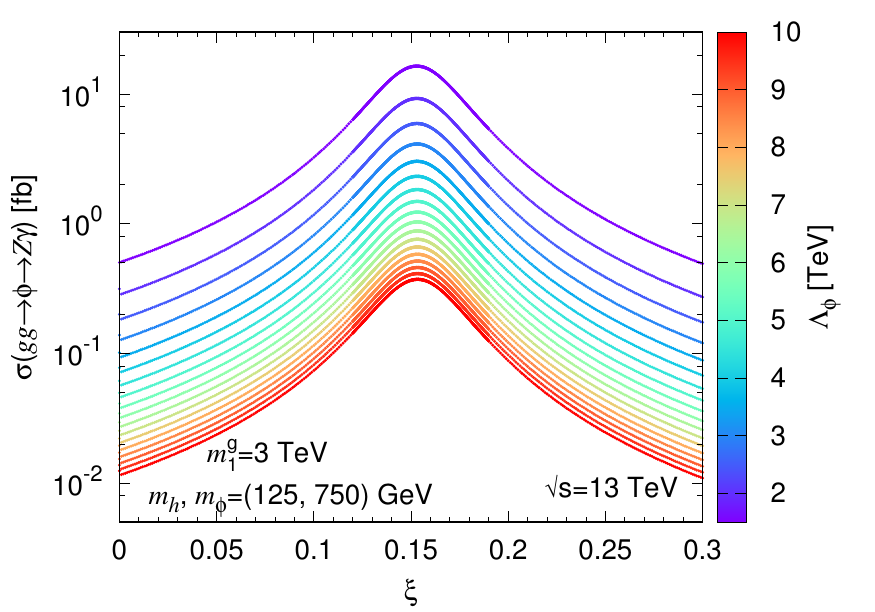}\\
\includegraphics[scale=0.62]{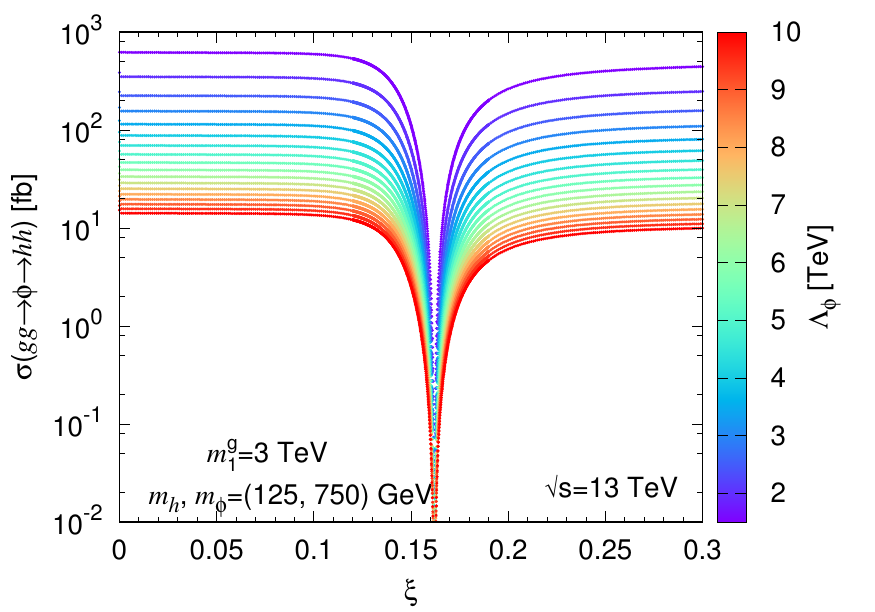}
\hspace{-1mm}\includegraphics[scale=0.62]{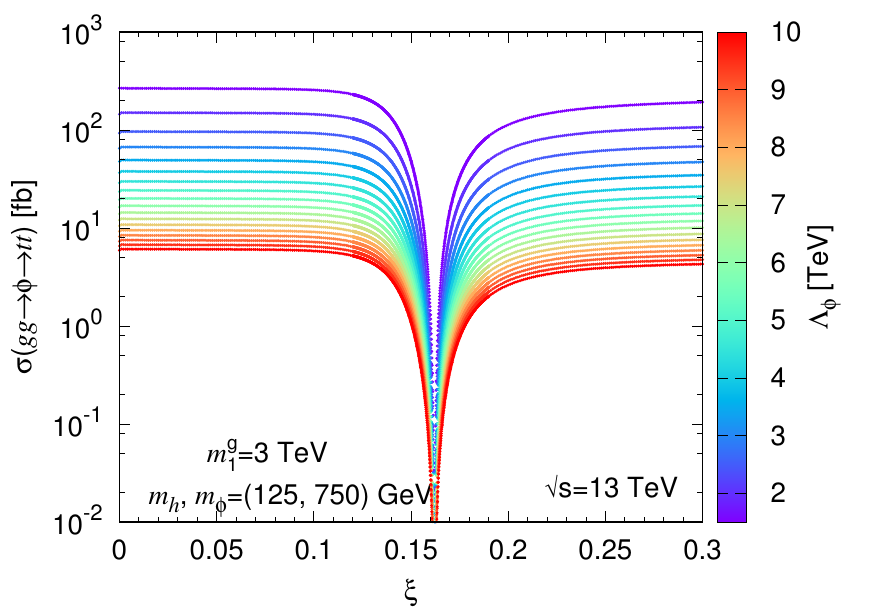}
\end{tabular}
\caption{We plot the cross-section $\sigma(gg\to\phi)\times \br(\phi\to X)$ for $X=WW,~ZZ,~Z\gamma,~hh$, $t\bar t$, as function of $\xi$ for $m_h=125\gev$, $m_\phi=750\gev$, $m_1^g=3\tev$ with different choices of $\Lambda_\phi$.}
\label{figphi3tev1}
\end{figure}

In Fig.~\ref{figphi3tev1} the cross-sections for $gg\to \phi$ with $\phi$ decay to $X=WW,~ZZ$, $Z\gamma$, $hh$ and $t\bar t$ final states, as functions of $\xi$, assuming $\mgone=3\tev$ are shown.
As can be seen from the plots in Fig.~\ref{figphi3tev1}, near the $g_\phi=0$ (\ie\ $\xi=0.162$) point  the $\sig(gg\to \phi)\times \br(\phi\to t\bar t,b\bar b,hh)$ values are highly suppressed, whereas the $\sig(gg\to \phi)\times \br(\phi\to WW,ZZ)$ are much less suppressed.
Of course, if $\mgone$ is larger, the cross sections for $WW,ZZ$ close to the $\xi$ value where $g_\phi=0$ will be smaller and somewhat more difficult to detect. (However, there is always a minimum value.)
Thus, in the vicinity of the dip region, the  $Z\gam$ final state would be the most relevant apart from $\gam\gam$ and $gg$. Indeed $\sig(gg\to \phi\to Z\gamma)$ is approximately twice as large as $\sig(gg\to \phi\to \gamma\gamma)$ at the resonance peak. Nonetheless, as we will quantify shortly, the $\gam\gam$ channel currently provides the stronger limit on $\lphi$.

\begin{figure}[t]
\centering
\begin{tabular}{ccc}
\hspace{-3mm}\includegraphics[width=0.5\textwidth]{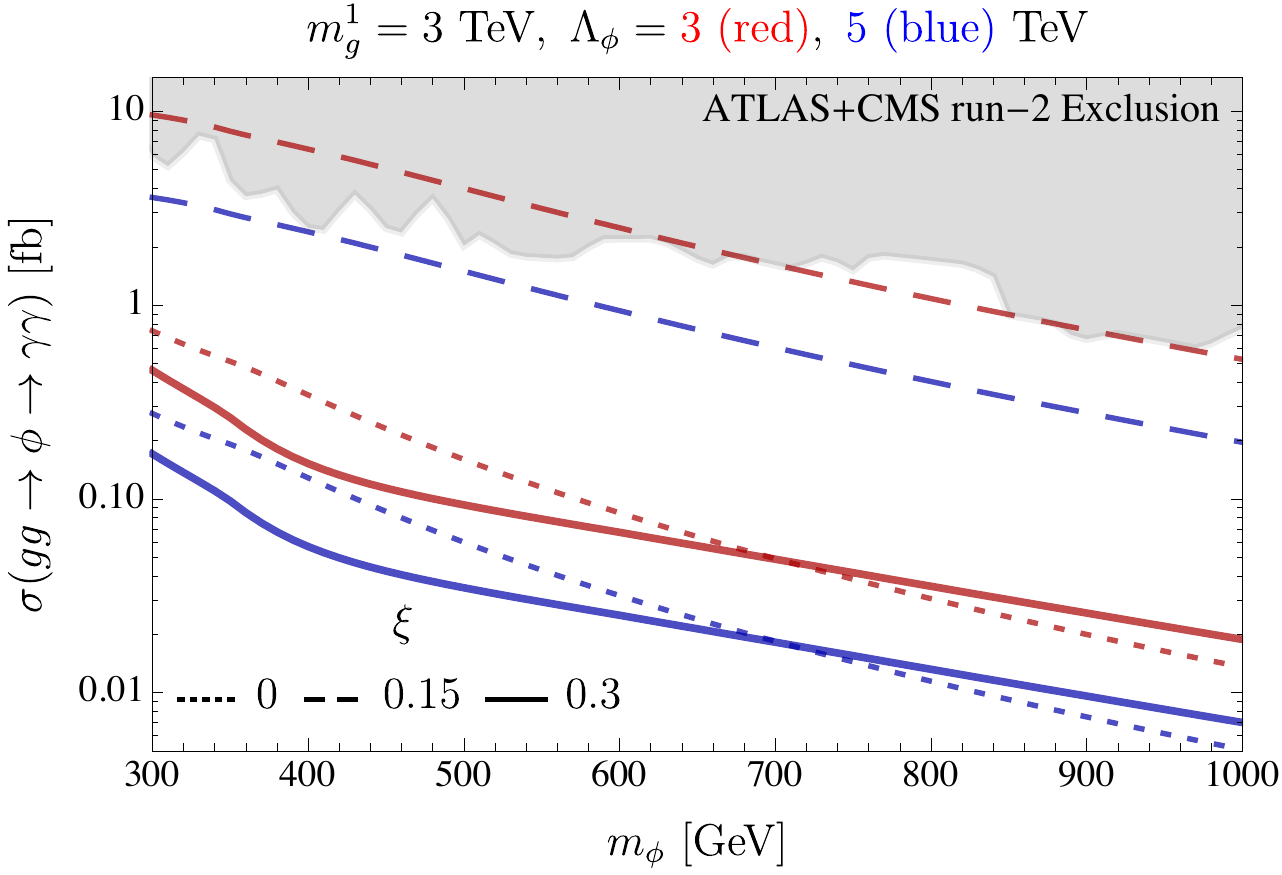}
\includegraphics[width=0.5\textwidth]{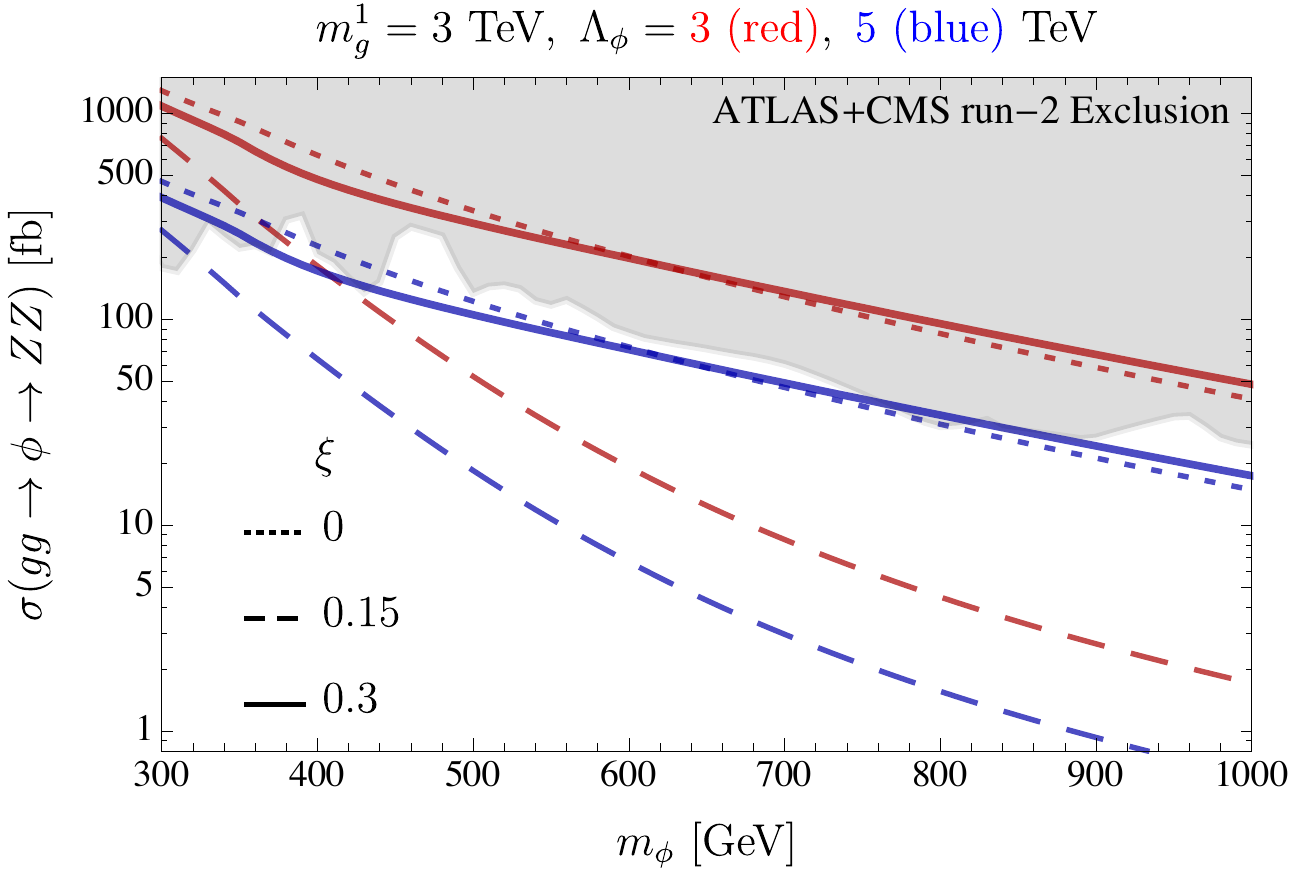}
\cr
\hspace{-3mm}\includegraphics[width=0.5\textwidth]{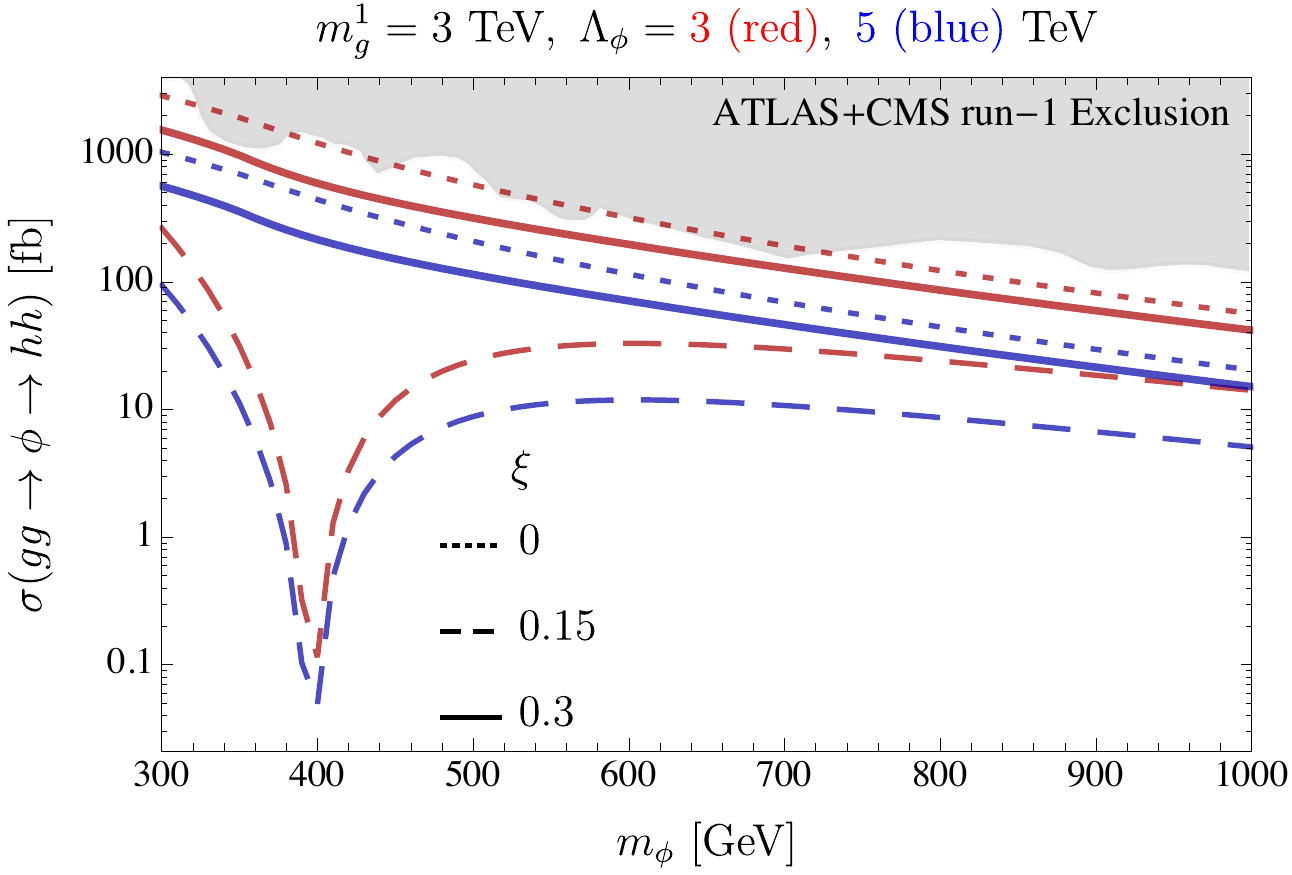}
\includegraphics[width=0.5\textwidth]{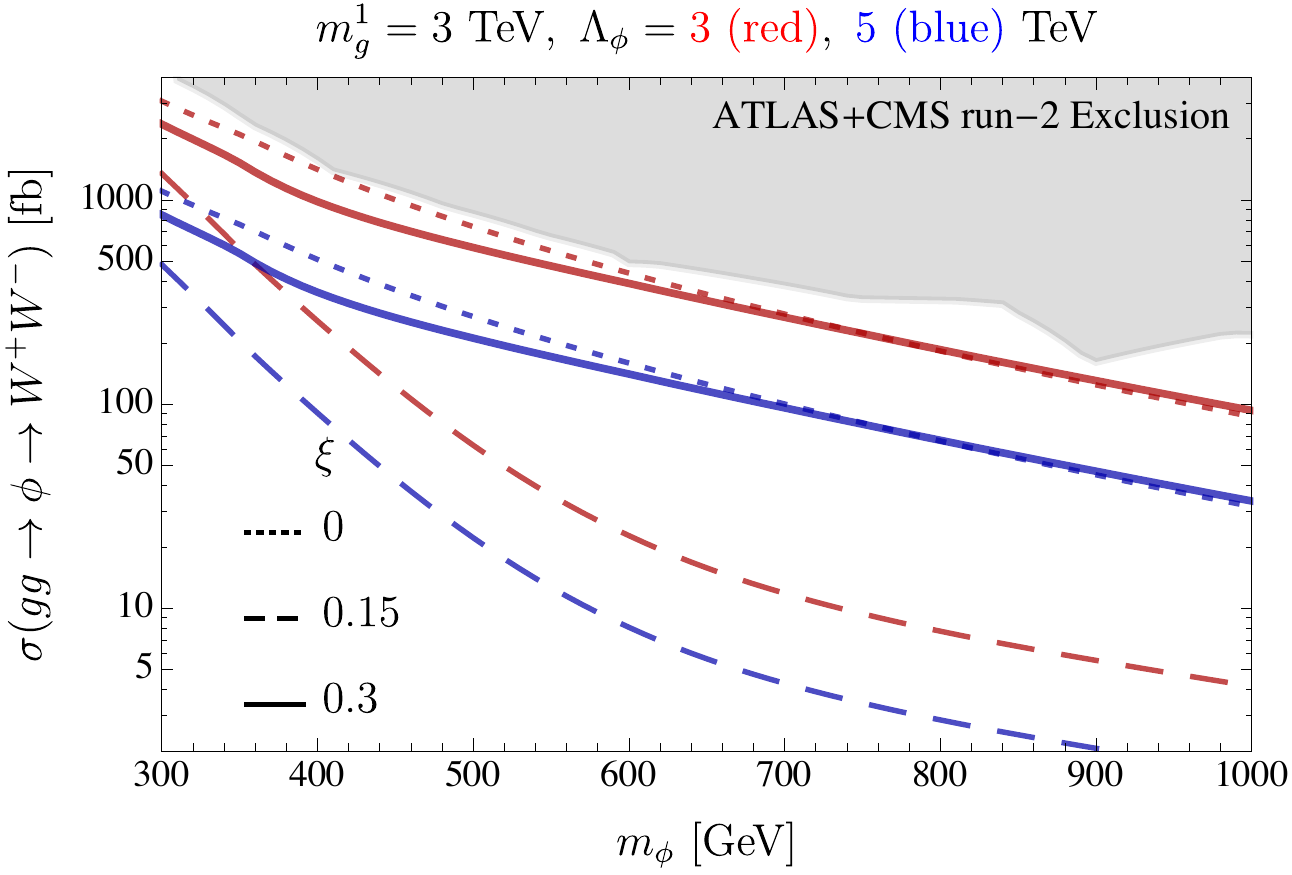}
\end{tabular}
\caption{ATLAS and CMS run-1 ($\rts\!=\!7/8\tev$) and run-2 ($\rts\!=\!13\tev$) limits on the $gg\to \phi\to X$ cross section for $X=\gam\gam,~ZZ,~hh$, and $WW$  as functions of $\mphi$ are shown by the grey regions in each of the respective plots. (The $\rts\!=\!8\tev$ limits are included by extrapolating to $\rts\!=\!13\tev$ using the standard $gg$ luminosity scaling factors provided by the HCWG~\cite{deFlorian:2016spz}.)   For those final states above which are actually observed in a variety of decay channels, we determined which decay channel led to the strongest limit after correcting for (\ie\ dividing by) the relevant branching ratio. At any given mass and for any one of the above final states, it is the strongest limit that is employed to place a limit on $\lphi$.(The complete list of experimental limits used is given in Appendix~\ref{expboundappendix}). 
The predictions of the model are shown for $\mgone\!=\!3\tev$ (red) and $\mgone\!=\!5\tev$ (blue) for the choices of  $\xi\!=\!0$, $0.15$ and $0.3$.}
\label{finalstatelimits}
\end{figure}

Of course, if $\xi$ is not near the region where the $\gam\gam,Z\gam,gg$ cross sections peak, then the strongest limits on $\lphi$ for a given $\mphi$ will come from the correspondingly unsuppressed  $ZZ,WW,hh,t\bar t$ final states. Thus, limits on all the final states must be incorporated in order to determine the minimum value of $\lphi$ for any given value of $\mphi$ (and choice of $\mgone$).  
The limits (see Appendix \ref{expboundappendix} for the experimental references employed) on the various final states as a function of resonance mass are plotted in Fig.~\ref{finalstatelimits} (the grey regions are excluded) in comparison to the predictions of the model for $\mgone=3\tev$ for several $\xi$ values.

As an example of how limits on $\lphi$ as a function of $\xi$ and $\mphi$ can be extracted from Fig.~\ref{finalstatelimits}, let us consider the case of $\mphi=750\gev$. Looking first at the $\xi=0.15$ predictions in the $\gam\gam$ final state for $\lphi=3$ and $5\tev$, we see that one enters into the excluded grey region for $\lphi\sim 3\tev$.  This should be the rough limit on $\lphi$ at $\mphi=750\gev$ near the conformal point, as will be plotted in the following figure. 
Next consider $\xi=0.3$, \ie\ far above the conformal point.  Looking now at the predictions (solid lines) in the $ZZ$ final state in comparison to the excluded grey region for the sample choice of $\mphi=750\gev$, we see that the prediction enters into the excluded region for a $\lphi$ value near $\sim 5\tev$, which is the limit that will appear in the coming figure. 

\subsection{Limits on $\lphi$}

\begin{figure}[h!]
\centering
\begin{tabular}{ccc}
\hspace{-5mm}\includegraphics[scale=0.95]{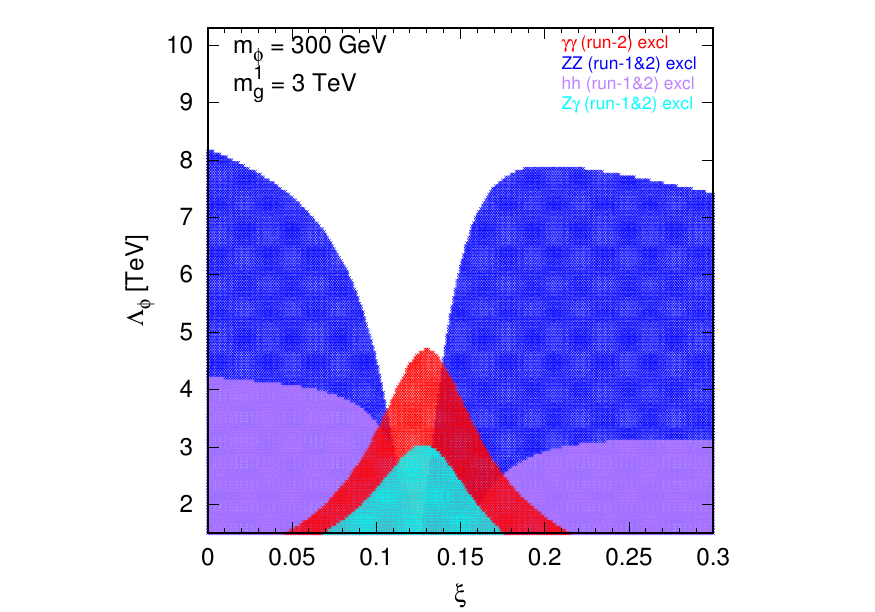}
\hspace{-15mm} \includegraphics[scale=0.95]{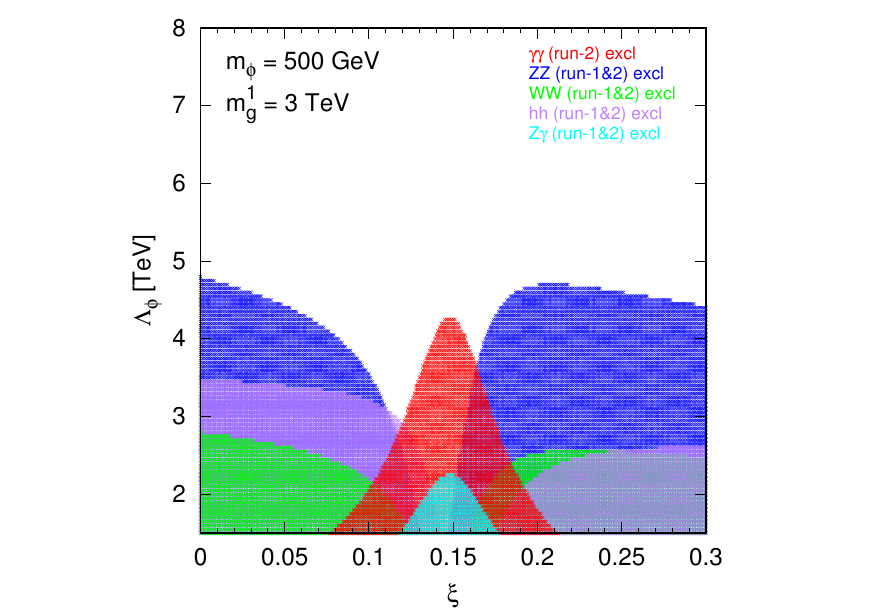}
\cr
\hspace{-5mm}\includegraphics[scale=0.95]{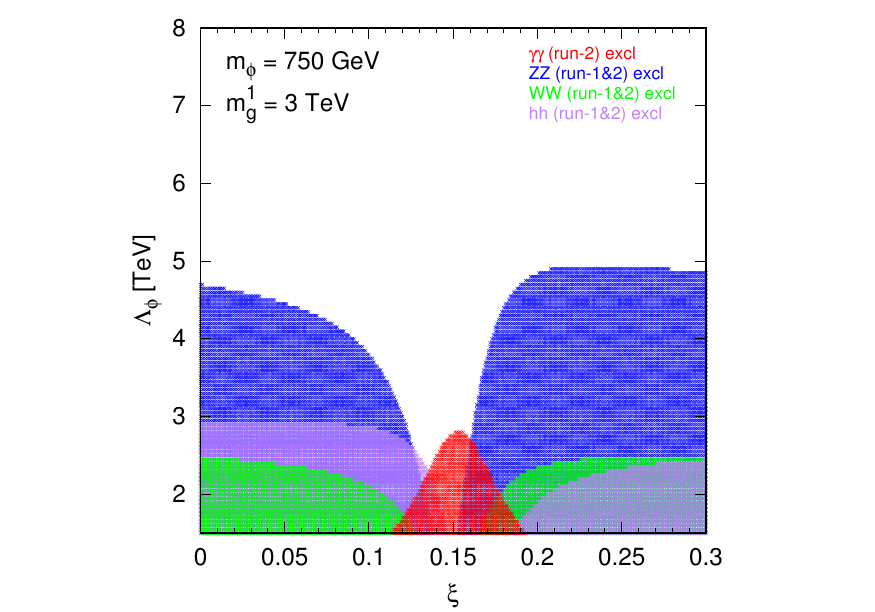}
\hspace{-15mm}\includegraphics[scale=0.95]{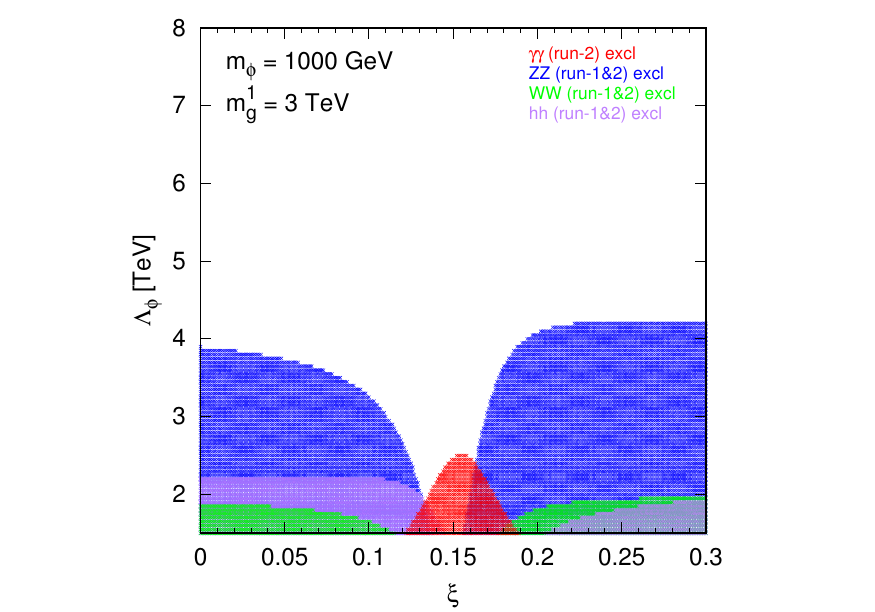}
\end{tabular}
\caption{We plot the limits on $\lphi$ coming from the important final states as functions of $\xi$ for $m_h=125\gev$, $m_\phi=750\gev$, $m_1^g=3\tev$.}
\label{figphi3tevexcl}
\end{figure}

The limits on $\lphi$ coming from the relevant final states as a function of $\xi$  for the sample cases of $\mphi=(300,500,750, 1000)\gev$ are  plotted in Fig.~\ref{figphi3tevexcl}. From these plots, we see that at the moment it is always the $\gam\gam$ final state that provides the strongest limit for $\xi$ values near the conformal point.  In contrast, well away from the conformal point,  it is the $ZZ$ final state that provides the strongest limit. The weakest bounds on $\lphi$ arise when $\xi$ is in the regions where the $\gam\gam$ cross section is declining and the other cross sections increasing.  These are roughly: $\lphi\gsim 3.8,2.9,2.4,2\tev$ for $\mphi=300,500,750,1000\gev$, respectively.

\begin{figure}[h]
\centering
\begin{tabular}{ccc}
\hspace{-6mm}\includegraphics[scale=0.59]{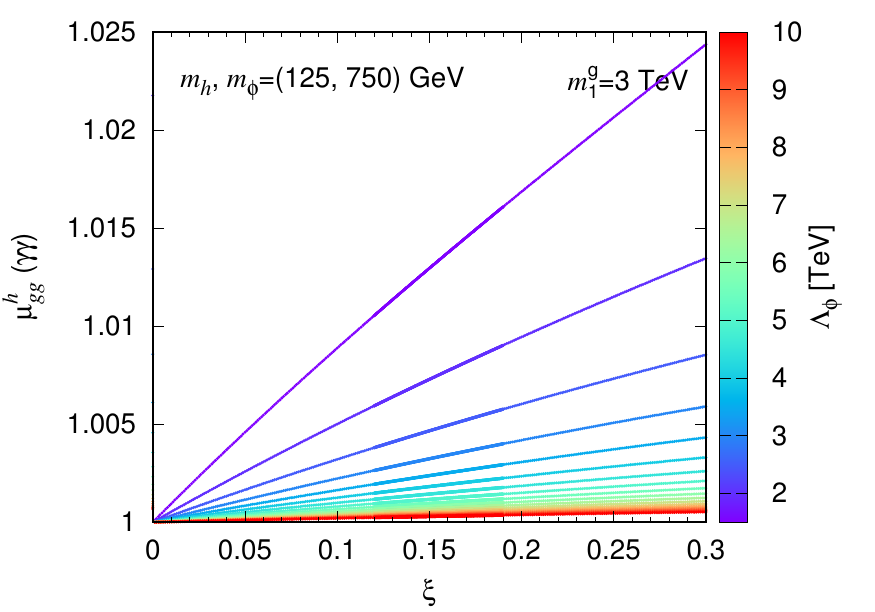}
\hspace{-1mm}\includegraphics[scale=0.59]{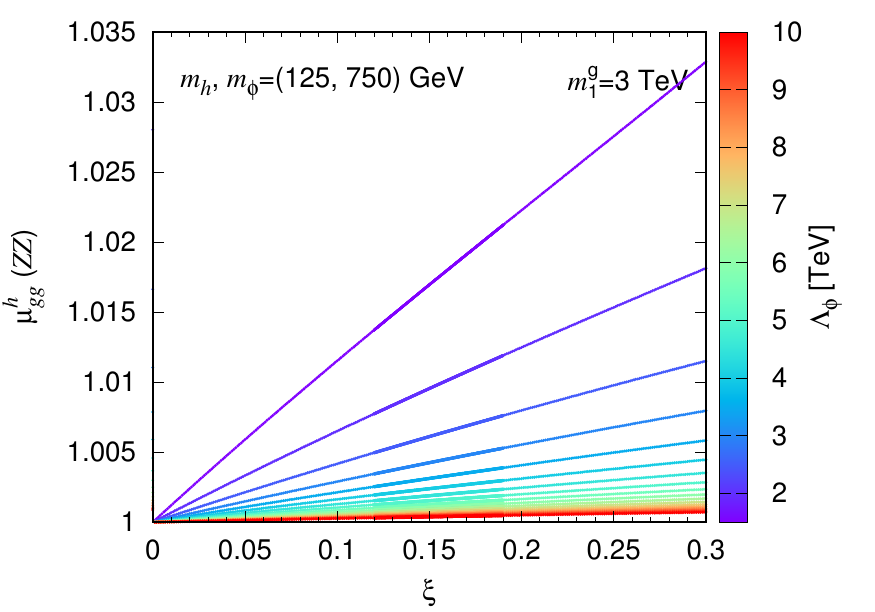}
\hspace{-1mm}\includegraphics[scale=0.59]{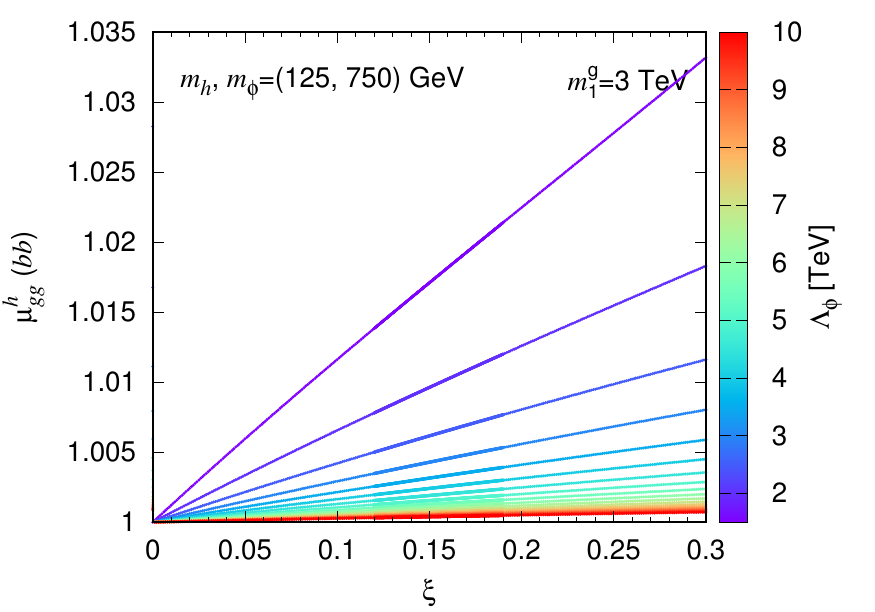}\\
\hspace{-6mm}\includegraphics[scale=0.59]{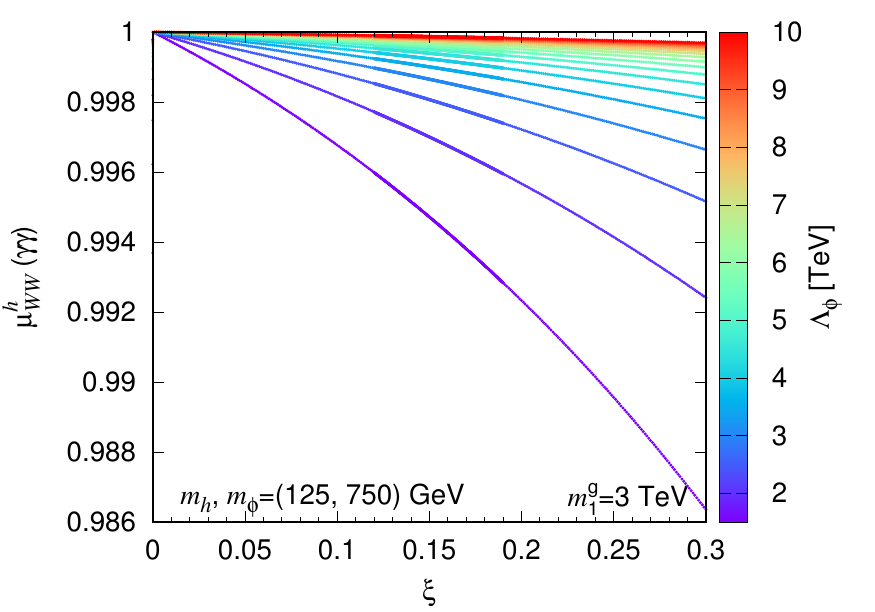}
\hspace{-1mm}\includegraphics[scale=0.59]{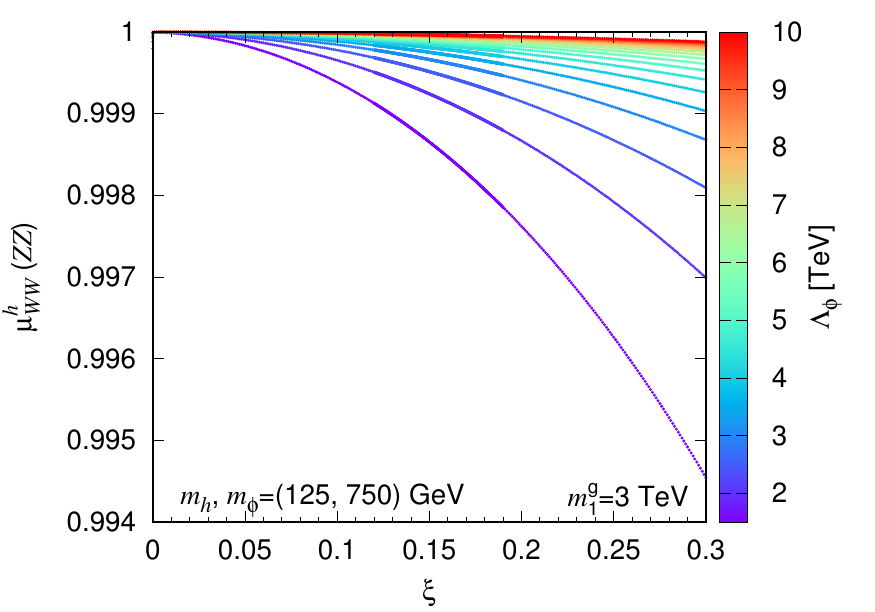}
\hspace{-1mm}\includegraphics[scale=0.59]{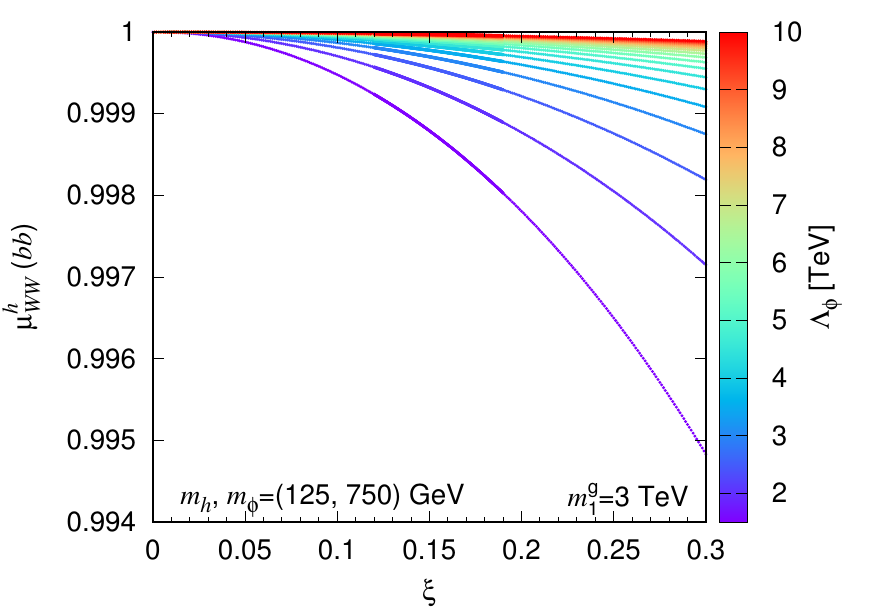}
\end{tabular}
\caption{The above graphs show $\mu_{gg,WW}^h(X)\equiv \sigma(gg,WW)\to h\to X)/\sigma_{\text{SM}}(gg,WW)\to h\to X)$, 
where $X=\gamma\gamma,~ZZ,~b\bar b$,  
as functions of $\xi$ for $m_h=125\gev$, $m_\phi=750\gev$, $m_1^g=3\tev$, color coded according to  $\Lambda_\phi$. 
}
\label{figrh3tev}
\end{figure}

Needless to say, for given choices of $\lphi$, $\mphi$ and $\mgone$ to be consistent it must also be the case that the Higgs-radion scenario  fit the Higgs data for the 125 GeV state. To this end, we examined the signal strength for all the measured channels and checked that corrections to the $125\gev$ couplings and production rates are all very small.  To illustrate, 
Figure \ref{figrh3tev} shows the ratios $\mu_{gg(WW)}^h(X)\equiv \sigma(gg(WW)\to h\to X)/\sigma_{\text{SM}}(gg(WW)\to h\to X)$, where $X=\gamma\gamma,~ZZ,~b\bar b$, as functions of $\xi$ for $m_h=125\gev$, $m_\phi=750\gev$, $m_1^g=3\tev$, color coded according to $\Lambda_\phi$. As expected, $\mu_{gg}^h(X) \simeq 1$ so that the $h$ is very SM-like for all $\xi$ values considered. The largest deviations from unity occur for the $gg$ induced processes, but are only of order a few percent even for the lowest $\lphi$ values we consider.~\footnote{We have not considered the KK-loop contributions to the Higgs and radion couplings to massless gauge bosons in our analysis. However, since in our work we employed the lower bound on the first KK-gluon mode as $m_1^g\geq 3\tev$, one can expect that the KK contributions to the loop-induced Higgs couplings to the massless gauge bosons would be at most $(10-15)\%$ relative to the $t$ and $W$ one-loop contributions, see for instance Ref.~\cite{Kubota:2012in}. On the other hand, the corresponding contribution to the couplings of the radion to massless gauge bosons is negligible as compared to the ``large'' anomalous and bulk contributions.}  However, if $\xi\gg0.3$ then the $\mu^h$ values will become sufficiently large that deviations from unity would become apparent with sufficiently large integrated luminosity.

\section{Conclusions}
\label{Conclusions}

In summary, the RS model is a solution to the hierarchy problem that does not require any additional symmetries or a plethora of new particles --- the only new particles would be those associated with the extra-dimensional context. The latter include the KK excitations of the gluon and other SM particles as well as the KK modes of the graviton and of the extra gauge bosons associated with the bulk custodial symmetry and, the focus of this paper,  the scalar radion associated with stabilizing the extra dimension. Thus, it is of great interest to explore limits on the model coming from the latest LHC data.

 In this work we have studied the phenomenology of a heavy radion  with mass $\mphi \geq 300\gev$ in the RS model.  A crucial feature in this model is the mixing between the radion  field and the Higgs field, introduced via a coupling of the Higgs field to the 4D Ricci scalar parameterized by the mixing parameter $\xi$.  The other important parameters of the model are the vacuum expectation value of the radion field, $\lphi$ and the mass of the first excited KK gluon, $\mgone$, Consistency of the model with limits on the brane curvature ratio ($k/\mpl<3$) imply that $\lphi$ values as low as $1\tev$ are possible if $\mgone=3\tev$ rising to $\lphi>1.75\tev$  if $\mgone=5\tev$ (see Fig.~\ref{figm1glamb}).  As a result, only for a very large limit on $\mgone$ will the model be inconsistent for values of $\lphi$ in the few TeV range.  Currently, direct searches and electroweak precision data are consistent with $\mgone$ as low as $3\tev$. Thus, it becomes relevant to determine the constraints on $\lphi$ in the RS model coming from direct searches for the radion.  Of course, these will depend on $\xi$ and $\mphi$ and, to a much lesser extent, on the value of $\mgone$.  In this paper, we have shown that limits on $\lphi$ are quite significant if the radion has $\mphi \geq 300\gev$.  (In a later paper, we explore scenarios with a light radion.)

The phenomenology of the radion-like eigenstate $\phi$ of the mixed Higgs-radion sector is quite sensitive to $\xi$.~\footnote{When $\mphi$ is substantially larger than $\mh\sim 125\gev$, the $h$ has very SM-like couplings unless $\xi$ is very large. Consequently, no constraints on the model arise from the $h$ eigenstate of the sector.} The conformal limit of the theory corresponds to $\xi=1/6$. Near the conformal point, all the couplings of the $\phi$ to the massive SM fields are suppressed while those to massless gauge bosons are large due to anomalous and bulk contributions, implying that at the LHC the $\phi$ would be primarily produced by $gg\to\phi$ and the branching ratios  $\br(\phi\to Z\gamma)$ and $\br(\phi\to\gamma\gamma)$ would be large. (For a large $gg\to \phi$ rate and for large $\gam\gam$ and $Z\gam$ branching ratios the bulk contributions to the corresponding couplings are especially important). In contrast, away from the conformal point the $\gam\gam$ and $Z\gam$ branching ratios are small while those to $WW,ZZ,t\bar t,hh$ are large.

Given that there is no sign of a resonance enhancement (of small or modest width) in any of these final states for any resonance mass, limits on the RS model as a function of $\lphi$, $\mphi$,  $\xi$, and $\mgone$  can be obtained.   We have found it convenient to summarize these limits as limits on $\lphi$ as a function for $\xi$ for given $\mgone$ and $\mphi$. Near the conformal point, the strongest limits on $\lphi$ derive from the $\gam\gam$ final state.  For $\xi$ values well away from the conformal point, the $ZZ$  final state typically provides the strongest limits. The weakest lower bound ranges from  $\lphi\gsim 3.8 \tev$ at $\mphi=300\gev$ to $\lphi\gsim 2 \tev$ at $\mphi=1\tev$ with very weak sensitivity to $\mgone$.
As the LHC continues operation, it remains entirely possible that a signal for the radion will be seen.  If not, the above bounds will  become substantially stronger and slowly decrease the attractiveness of a TeV scale RS solution to the hierarchy problem. Assuming no improvements in efficiencies and analysis techniques, since the radion cross sections scale as $1/\lphi^2$, the limits on $\lphi$ will scale with integrated luminosity as $\lphi^{\text{limit}}\propto [L_{\text{int}}]^{1/4}$, for fixed $\rts=13\tev$.

\section*{Acknowledgements}

The work of AA and BG has been supported in part by the National Science Centre
(Poland) as  research project no DEC-2014/13/B/ST2/03969. BMD was supported by the Science and Technology Facilities Council (UK). JFG is supported in part by the US DOE grant DE-SC-000999. BMD and JFG thank the University of Warsaw for hospitality.
YJ acknowledges generous support by the Villum Foundation. AA is grateful to the Mainz Institute for Theoretical Physics (MITP) for hospitality and support during the completion of this work.

\newpage
\appendix
\section{Experimental bounds}
\label{expboundappendix}
In this Appendix, Table~\ref{exptBounds} collects all the experimental ATLAS and CMS references for Run-1 and Run-2 employed in our analysis.
\begin{table}[h!]
\centering
\caption{Relevant experimental data from ATLAS and CMS experiments for Run-1 and Run-2 employed in our analysis.}
\label{exptBounds}
\vspace{-13pt}
\begin{center}
{\small
{\tabulinesep=3pt
\rowcolors{1}{gray!10}{white}
\begin{tabular}{|l|l|l|c|l|}
\hline
Experiment           &Final state decay channels & Mass range & Reference  &  Notes   \\
 \hline
 \hline
ATLAS Run-1 &      $ZZ\rightarrow llll+ll\nu\nu+llqq+\nu\nu qq$         &   $140-1000$ GeV         &  \cite{Aad:2015kna}   &  $20.3\text{ fb}^{-1}$   \\
           &      $WW\rightarrow l\nu l\nu+l\nu qq$         &   $300-1500$ GeV         &   \cite{Aad:2015agg}   &  $20.3\text{ fb}^{-1}$   \\
           &         $\gamma\gamma$         &     $500-3000$ GeV       &  \cite{Aad:2015mna}   &  $20.3\text{ fb}^{-1}$    \\
           &        $hh\rightarrow b\bar{b}b\bar{b}$         &    $500-1500$ GeV        &  \cite{ATLAS:2014rxa}    & $19.5\text{ fb}^{-1}$   \\
           &        $hh\rightarrow b\bar{b}\gamma\gamma$         &    $260-500$ GeV        &  \cite{Aad:2014yja}    & $20\text{ fb}^{-1}$   \\
           &           $\bar{t}t$          &   $400-2600$ GeV         &  \cite{Aad:2015fna}   &  $20.3\text{ fb}^{-1}$   \\
                      &       $\tau\tau\rightarrow$lep+had+lep/had         &   $94-1000$ GeV         &  \cite{Aad:2014vgg}    &  $20.3\text{ fb}^{-1}$  \\
           &        $Z\gamma$         &     $300-1600$ GeV       &  \cite{Aad:2014fha}     & $20.3\text{ fb}^{-1}$  \\
           &       $jj$         &  $400-4200$ GeV          &  \cite{Aad:2014aqa}    &  $20.3\text{ fb}^{-1}$  \\
\hline
CMS Run-1           &      $\gamma\gamma$       &     $150-900$ GeV       &   \cite{CMS:2014onr}  &  $19.7\text{ fb}^{-1}$   \\
           &          $hh\rightarrow b\bar{b}\gamma\gamma$          &     $260-1100$ GeV       &  \cite{Khachatryan:2016sey}    & $19.7\text{ fb}^{-1}$   \\
           &           $hh\rightarrow b\bar{b}b\bar{b}$          &     $260-1100$ GeV       &  \cite{Khachatryan:2015yea,Khachatryan:2016cfa}    & $17.9\text{ fb}^{-1}$   \\
           &           $hh\rightarrow b\bar{b}\tau\tau$          &     $230-2700$ GeV       &  \cite{Khachatryan:2015tha,CMS:2016zxv,CMS:2015zug}    & $18.3\text{ fb}^{-1}$   \\
\hline
\hline
ATLAS Run-2 &    $ZZ\rightarrow llll$           &   $200-1000$ GeV         &   \cite{ATLAS:2016oum}  & $14.8\text{ fb}^{-1}$    \\
                     &    $ZZ\rightarrow ll\nu\nu$           &   $300-1000$ GeV         &   \cite{ATLAS:2016bza}  & $13.3\text{ fb}^{-1}$    \\
           &    $ZZ\rightarrow llqq$           &   $330-3000$ GeV         &   \cite{ATLAS:2016npe}  & $13.2\text{ fb}^{-1}$    \\
           &    $ZZ\rightarrow \nu\nu qq$           &   $350-3000$ GeV         &   \cite{ATLAS:2016npe}  & $13.2\text{ fb}^{-1}$    \\
           &      $WW\rightarrow e\nu\mu\nu$          &      $300-3000$ GeV      &  \cite{ATLAS:2016kjy}    & $13.2\text{ fb}^{-1}$   \\
           &       $WW\rightarrow l\nu qq$          &    $500-3000$ GeV        &  \cite{ATLAS:2016cwq}    &  $13.2\text{ fb}^{-1}$  \\
           &        $\gamma\gamma$          &      $200-2400$ GeV      & \cite{ATLAS:2016eeo}    & $15.4\text{ fb}^{-1}$    \\
           &        $hh$         &  $260-500$ GeV          &   \cite{ATLAS:2016qmt}   & $13.3\text{ fb}^{-1}$   \\
           &           $\bar{t}t$          &   $400-1000$ GeV         &  \cite{ATLAS:2016btu}   &  $13.2\text{ fb}^{-1}$   \\
           &      $\tau\tau\rightarrow$ had+lap/had          &    $200-1200$ GeV        &  \cite{Aaboud:2016cre}    &  $3.2\text{ fb}^{-1}$   \\
           &     $Z\gamma\rightarrow \stackrel{<1500\text{ GeV}}{ll\gam}+\stackrel{>700\text{ GeV}}{qq\gam}$          &            $260-2750$ GeV       &   \cite{Aaboud:2016trl}    & $3.2\text{ fb}^{-1}$ \\
           &        $Z\gam\rightarrow ll\gam$       &    $250-2400$ GeV        &   \cite{ATLAS:2016lri} &  $3.2\text{ fb}^{-1}$    \\
\hline
CMS Run-2   &   $ZZ\rightarrow llll$           &    $140-2500$ GeV        &  \cite{CMS:2016ilx}   & $12.9\text{ fb}^{-1}$    \\
           &       $WW\rightarrow l\nu l\nu$          &     $200-1000$ GeV       &  \cite{CMS:2016jpd}    & $2.3\text{ fb}^{-1}$   \\
        &    $\gamma\gamma$ (combined Run-1/2)            &     $500-3500$ GeV       &  \cite{Khachatryan:2016yec}   &  $3.3$/$19.7\text{ fb}^{-1}$    \\
           &           $hh\rightarrow b\bar{b}b\bar{b}$          &    $230-1200$ GeV        &   \cite{CMS:2016tlj}   & $2.3\text{ fb}^{-1}$   \\
           &           $hh\rightarrow b\bar{b}b\bar{b}$          &     $900-3300$ GeV       &  \cite{CMS:2016pwo}    & $2.7\text{ fb}^{-1}$   \\
           &           $hh\rightarrow b\bar{b}\gam\gam$          &     $250-900$ GeV       &   \cite{CMS:2016vpz}   &  $2.7\text{ fb}^{-1}$  \\
           &           $hh\rightarrow b\bar{b}\tau\tau$          &    $250-900$ GeV        &  \cite{CMS:2016knm}    & $12.9\text{ fb}^{-1}$   \\
           &      $hh\rightarrow b\bar{b}l\nu l\nu$          &    $260-500$ GeV        &   \cite{CMS:2016rec}   & $2.3\text{ fb}^{-1}$   \\
           &     $\tau\tau\rightarrow$ lep+had+lep/had          &    $90-3200$ GeV        &   \cite{CMS:2016rjp}   &  $12.9\text{ fb}^{-1}$  \\
           &     $\tau\tau\rightarrow$ lep+had+lep/had          &    $90-3200$ GeV        &  \cite{CMS:2016rjp}    &  $12.9\text{ fb}^{-1}$  \\
           &       $bb$         &  $550-1200$ GeV          & \cite{CMS:2016ncz}  & $2.69\text{ fb}^{-1}$ \\
\hline
\end{tabular}
}}
\end{center}
\end{table}
\newpage

\providecommand{\href}[2]{#2}\begingroup\raggedright\endgroup

\end{document}